\documentclass{article}
\usepackage{arxiv}
\usepackage[T1]{fontenc}
\usepackage[utf8]{inputenc}
\usepackage[english]{babel}
\usepackage{lmodern}
\usepackage{graphicx}
\usepackage{amsmath,amssymb,amsfonts}
\usepackage{bm}
\usepackage{dsfont}
\usepackage{upgreek}
\usepackage{booktabs}
\usepackage{multirow}
\usepackage{array}
\usepackage{rotating}
\usepackage{pdflscape}
\usepackage{placeins}
\usepackage{longtable}
\usepackage[para,online,flushleft]{threeparttable}
\usepackage{floatrow}
\newfloatcommand{capbtabbox}{table}[][\FBwidth]
\usepackage{subcaption}
\usepackage[style=base,figurename=Fig.,labelfont=bf,labelsep=period]{caption}
\usepackage[square,sort&compress,numbers]{natbib}
\usepackage{doi}
\usepackage{theorem}
\usepackage{etoolbox}
\usepackage{orcidlink}
\usepackage{newtxtext,newtxmath}
\usepackage{tikz}
\definecolor{blendedblue}{rgb}{0.2, 0.2, 0.6}
\usepackage{hyperref}
\hypersetup{
  colorlinks  = true,
  linkcolor   = blendedblue,
  citecolor   = blendedblue,
  urlcolor    = blendedblue,
  linktocpage = false
}
\usepackage{siunitx}
\newcolumntype{Y}{>{\RaggedRight\arraybackslash}X}
\newcolumntype{Z}[1]{>{\RaggedRight\arraybackslash}p{#1}}

\newcolumntype{L}[1]{>{\RaggedRight\arraybackslash}p{#1}}
\newcolumntype{O}[1]{>{\centering\arraybackslash}p{#1}}

\newcolumntype{A}{>{\RaggedRight\arraybackslash}p{0.20\textwidth}}
\newcolumntype{B}{>{\RaggedRight\arraybackslash}p{0.115\textwidth}}
\newcolumntype{C}{>{\RaggedRight\arraybackslash}X}
\newcolumntype{D}{>{\RaggedRight\arraybackslash}p{0.145\textwidth}}
\newcolumntype{E}{>{\RaggedRight\arraybackslash}p{0.155\textwidth}}
\newcolumntype{T}{r}
\newcolumntype{N}{S[table-format=2.0]}

\usepackage{ragged2e}
\usepackage{tabularx}
\usepackage[normalem]{ulem}

\begin{document}
\title{Data Set of Load Tests and Structural Health Monitoring of a concrete Box Girder bridge}

\title{Data Set of Load Tests and Structural Health Monitoring of a concrete boxgirder bridge}
\author{
        Martin Köhncke\orcidlink{0009-0000-7942-3994}\\
 	\footnotesize{Chair of Engineering Materials and Building Preservation}\\
	\footnotesize{Faculty of Mechanical and Civil Engineering}\\
        \footnotesize{Helmut Schmidt University}\\
	\footnotesize{Hamburg, Germany}\\
	\footnotesize{\texttt{martin.koehncke@hsu-hh.de}} \\
        \And
        Yogi Jaelani\orcidlink{0009-0005-0297-7426}\\
 	\footnotesize{Chair of Engineering Materials and Building Preservation}\\
	\footnotesize{Faculty of Mechanical and Civil Engineering}\\
        \footnotesize{Helmut Schmidt University}\\
	\footnotesize{Hamburg, Germany}\\
        \And
        Alexander Mendler\orcidlink{0000-0002-7492-6194}\\
        \footnotesize{Dept. of Materials Engineering}\\
        \footnotesize{TUM School of Engineering and Design}\\
	\footnotesize{Technical University of Munich}\\
        \footnotesize{Munich, Germany}\\
        \And
        Lizzie Neumann\orcidlink{0000-0003-2256-1127}\\
 	\footnotesize{Dept. of Mathematics and Statistics}\\
	\footnotesize{School of Economics and Social Sciences}\\
        \footnotesize{Helmut Schmidt University}\\
	\footnotesize{Hamburg, Germany}\\
  	\And
        Philipp Wittenberg\orcidlink{0000-0001-7151-8243}\\
 	\footnotesize{Dept. of Mathematics and Statistics}\\
	\footnotesize{School of Economics and Social Sciences}\\
        \footnotesize{Helmut Schmidt University}\\
	\footnotesize{Hamburg, Germany}\\
        \And
        Alina Rode-Klemm\orcidlink{0009-0008-7375-9460}\\
 	\footnotesize{Dept. of Steel Structures}\\
	\footnotesize{Faculty of Mechanical and Civil Engineering}\\
        \footnotesize{Helmut Schmidt University}\\
	\footnotesize{Hamburg, Germany}\\
        \And
        Sylvia Kessler \orcidlink{0000-0002-1335-1104}\\
 	\footnotesize{Chair of Engineering Materials and Building Preservation}\\
	\footnotesize{Faculty of Mechanical and Civil Engineering}\\
        \footnotesize{Helmut Schmidt University}\\
	\footnotesize{Hamburg, Germany}\\  
}

\newcommand\copyrighttext{%
  \scriptsize This manuscript has been accepted for publication in the \textit{Journal of Bridge Engineering}. The manuscript will be updated with the complete citation and DOI after publication. \textcopyright{} 2026 The Authors. This work is licensed under the
\href{https://creativecommons.org/licenses/by/4.0/}
{Creative Commons Attribution 4.0 International License}.}
\newcommand\copyrightnotice{%
\begin{tikzpicture}[remember picture,overlay]
\node[anchor=south,yshift=10pt] at (current page.south) {\fbox{\parbox{\dimexpr\textwidth-\fboxsep-\fboxrule\relax}{\copyrighttext}}};
\end{tikzpicture}%
}

\date{ }
\maketitle
%%%%%%%%%%%%%%%%%%%%%%%%%%%%%%%%%%%%%%%%%%%%%%%%
\copyrightnotice
\begin{abstract}
Static and dynamic load tests were conducted on an anonymized in-service prestressed concrete box girder bridge constructed in 1972 and designed for Bridge Class~30 according to DIN~1072.
The bridge is instrumented as part of the DTEC-SHM research initiative with a permanent structural health monitoring system comprising 128 sensors, including accelerometers, strain sensors, inclinometers, displacement transducers, embedded temperature sensors, and weather stations.
The test program includes long-term static loading with stepwise added masses of \SI{680}{kg}, \SI{1420}{kg}, and \SI{2160}{kg}, short-term static truck parking positions, and repeated dynamic truck crossings under loaded and unloaded vehicle configurations.
The short-term static tests used a loaded truck of \SI{21250}{kg} parked at three bridge positions, while the dynamic tests used loaded and unloaded truck configurations of \SI{21250}{kg} and \SI{12340}{kg} at speeds up to \SI{13.9}{m/s}.
The published load-test data are available on Zenodo in HDF5 format and include synchronized measurements from the monitoring system, temporary laser sensors used for vehicle-position and passage-time validation, and derived tracked eigenfrequency time series from the reference and long-term static-load-test period.
The bridge geometry, sensor layout, acquisition system, load cases, truck geometry, and data hierarchy are documented to support reuse of the data.
Plausibility checks are presented for the main response quantities, including laser-based validation of parking positions and passage windows, vibration spectra and tracked modal parameters, and repeatable strain and displacement response profiles under crawl-speed truck passages.
The data set addresses the scarcity of publicly available full-scale bridge load-test data and supports applications such as finite element model calibration, neutral-axis tracking, influence-line extraction, and data-driven anomaly detection.
\end{abstract}

%%%%%%%%%%%%%%%%%%%%%%%%%%%%%%%%%%%%%%%%%%%%%%%%
\section{Introduction}
%%%%%%%%%%%%%%%%%%%%%%%%%%%%%%%%%%%%%%%%%%%%%%%%

Load tests are essential in bridge engineering to ensure the safety and longevity of both new and existing bridges.
They are used to validate design assumptions, calibrate finite element models, assess vibration behavior, and identify potential structural weaknesses. 
In the case of existing bridges, load tests provide essential information on actual in-service performance, structural stiffness, fatigue behavior, and remaining load capacity, thereby supporting condition assessment \citep{Faber.etal_2000, Lantsoght.etal_2017, Lantsoght_2022}.
Two primary types of load tests are commonly distinguished: Static load tests measure the bridge's response to loads applied gradually or maintained for a certain period of time, while dynamic load tests assess the response to loads that rapidly change over time, such as moving trucks.
During static load tests, a stationary load is applied, often in increments, until the structure’s static response can be measured (e.g., displacements, inclinations, deflections, strains). 
These can be used for damage detection, such as monitoring confounder-adjusted scores or signals~\citep{Neumann_2025, Neumann.etal_2026b}. 
Furthermore, the added masses (up to \SI{2160}{kg}) correspond to only about \SIrange{0.04}{0.13}{\percent} of the estimated bridge mass of \SI{1670000}{kg}; hence, the sensitivity of monitoring methods can be tested. The bridge mass was determined approximately based on the information in the plans.
Dynamic load tests focus on measuring the bridge’s dynamic response (accelerations, velocities, displacements) to time-varying loads, such as moving trains, trucks, or seismic loads. 
They can be used for Bayesian updating of finite element models~\citep{Marsili.etal_2017, Lee.Cho_2016} or predictive probability of detection (P-POD) curves~\citep{Mendler.etal_2024}. 

This paper describes the setting and execution of static and dynamic load tests on a concrete box girder bridge.  As part of a greater research initiative (DTEC-SHM), the Helmut Schmidt University in Hamburg (HSU) has installed a permanent monitoring system comprising more than one hundred sensors. The objectives of this paper are to (a) document the monitoring system, (b) conduct and document both static and dynamic load tests to capture the current condition state for future reference under various loads, and examine the stress distribution and structural response to identify potential weaknesses, (c) record measurement data with minimal mass variations to assess the monitoring system's sensitivity and (d) record measurement data for model calibration and validating anomaly detection algorithms by simulating one design load case and (e) providing measurement data from full-scale load tests to enable further scientific research. To date, there are only a few open-access datasets from different bridge types with SHM systems for scientific research. This publication aims to address this shortcoming of the infrequent occurrence of such load tests and the limited availability of data from these experiments identified by \cite{Whelan.etal_2025}.  
In addition, the data can also be utilized for the extraction of specific features, such as the neutral axis location \citep{Sigurdardottir.etal_2013} or influence lines \citep{Zheng.etal_2021}. A shift in the neutral axis for a given cross-section may indicate unusual structural behavior. Even simple monitoring systems with two parallel sensors and repeated short-term measurements with large strain magnitudes provide reliable results. However, \cite{Sigurdardottir.etal_2013} point out that further research is needed on the influence of temperature gradients and on other bridge types. \cite{Soman.Ostachowicz_2018} used a robust version of the Kalman Filter to use neutral axis locations as a damage-sensitive feature. Influence lines can be used to identify (structural) damage or to support transfer learning \citep{Ferguson.etal_2025}. \cite{Huang.Lu_2025} presented a single-sensor M-value feature based on strain gauges to identify structural damage. This was explained using influence lines and the variation of the feature in connection to shifts in the neutral axis position. A comprehensive review of different methods and the use of influence lines is given by \cite{Cai.etal_2025}. 
The scientific value of the present contribution does not lie in an unusual or damaged bridge structure, but in the controlled combination of a full-scale in-service bridge, dense synchronized instrumentation, documented static and dynamic load cases, independent laser-based vehicle-position measurements, and openly provided load-test data.
This combination enables researchers to test preprocessing, feature extraction, model calibration, and anomaly-detection workflows on a realistic bridge response where load magnitudes, vehicle geometry, vehicle timing, and environmental measurements are documented together.

%%%%%%%%%%%%%%%%%%%%%%%%%%%%%%%%%%%%%%%%%%%%%%%%
\section{Bridge Structure}\label{sec:bridge}
%%%%%%%%%%%%%%%%%%%%%%%%%%%%%%%%%%%%%%%%%%%%%%%%

\begin{figure}[!htb]
    \centering
    \includegraphics[height = 4.1cm]{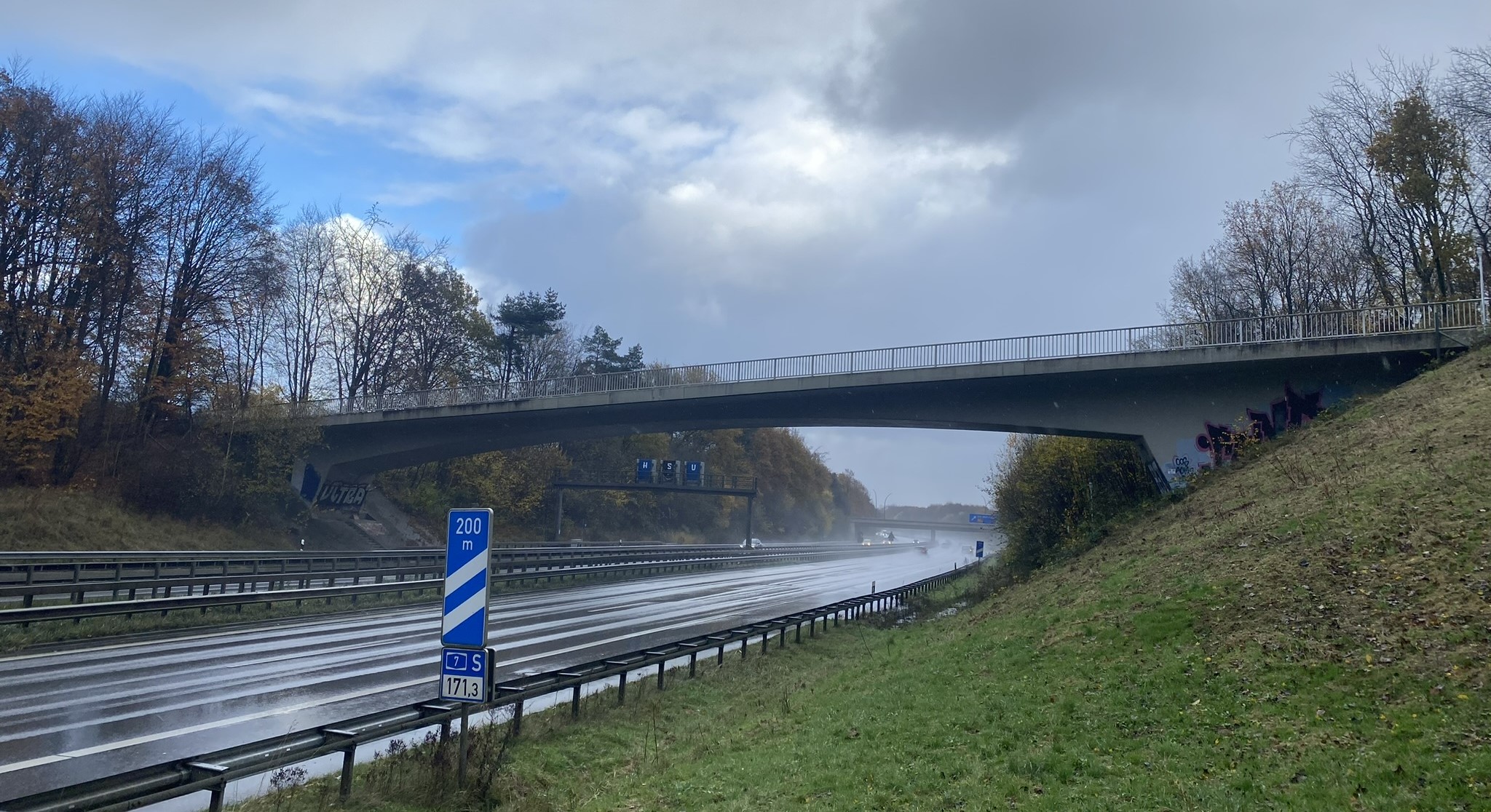}
    \includegraphics[height = 4.1cm]{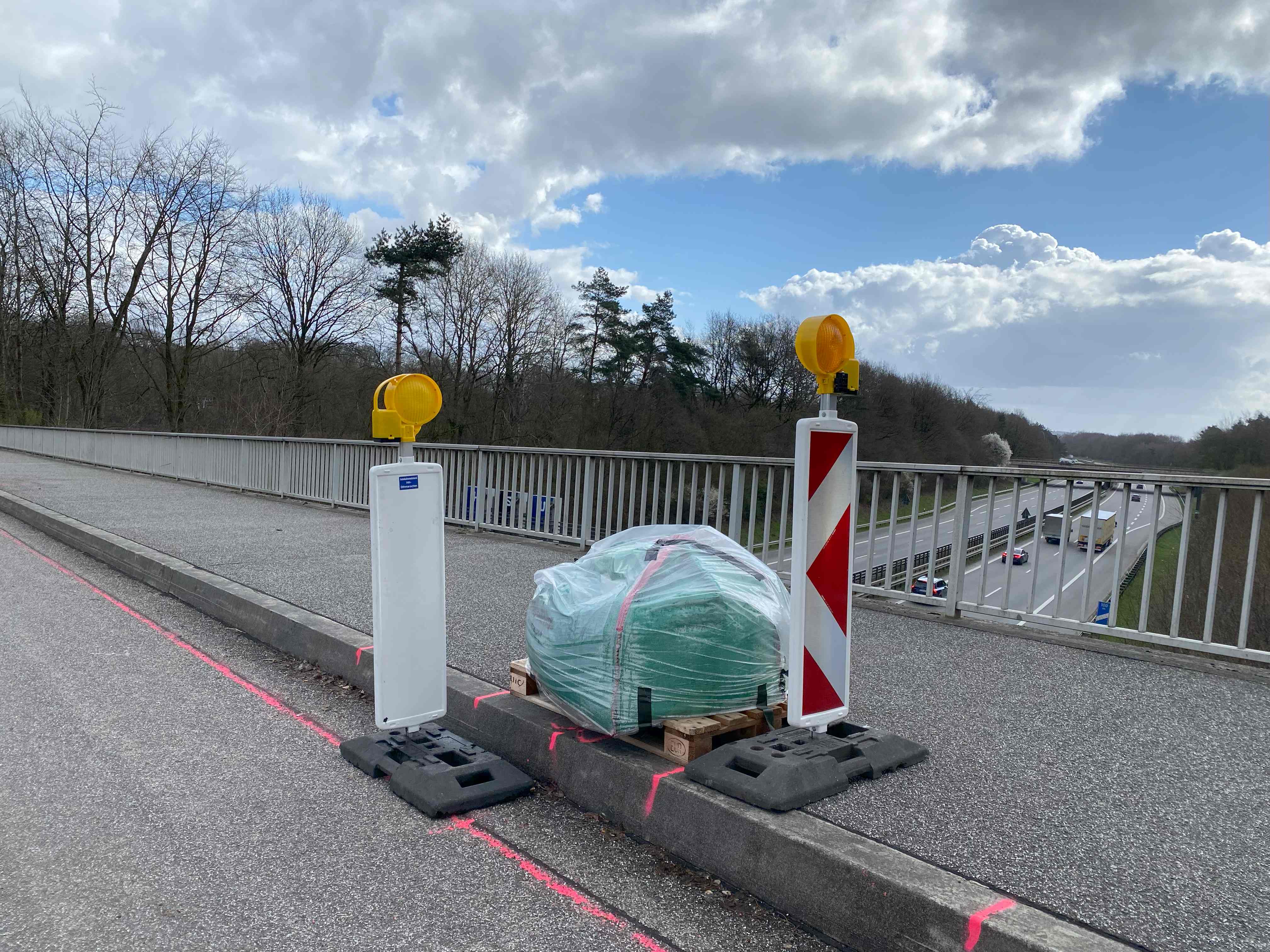}
    \includegraphics[height = 4.1cm]{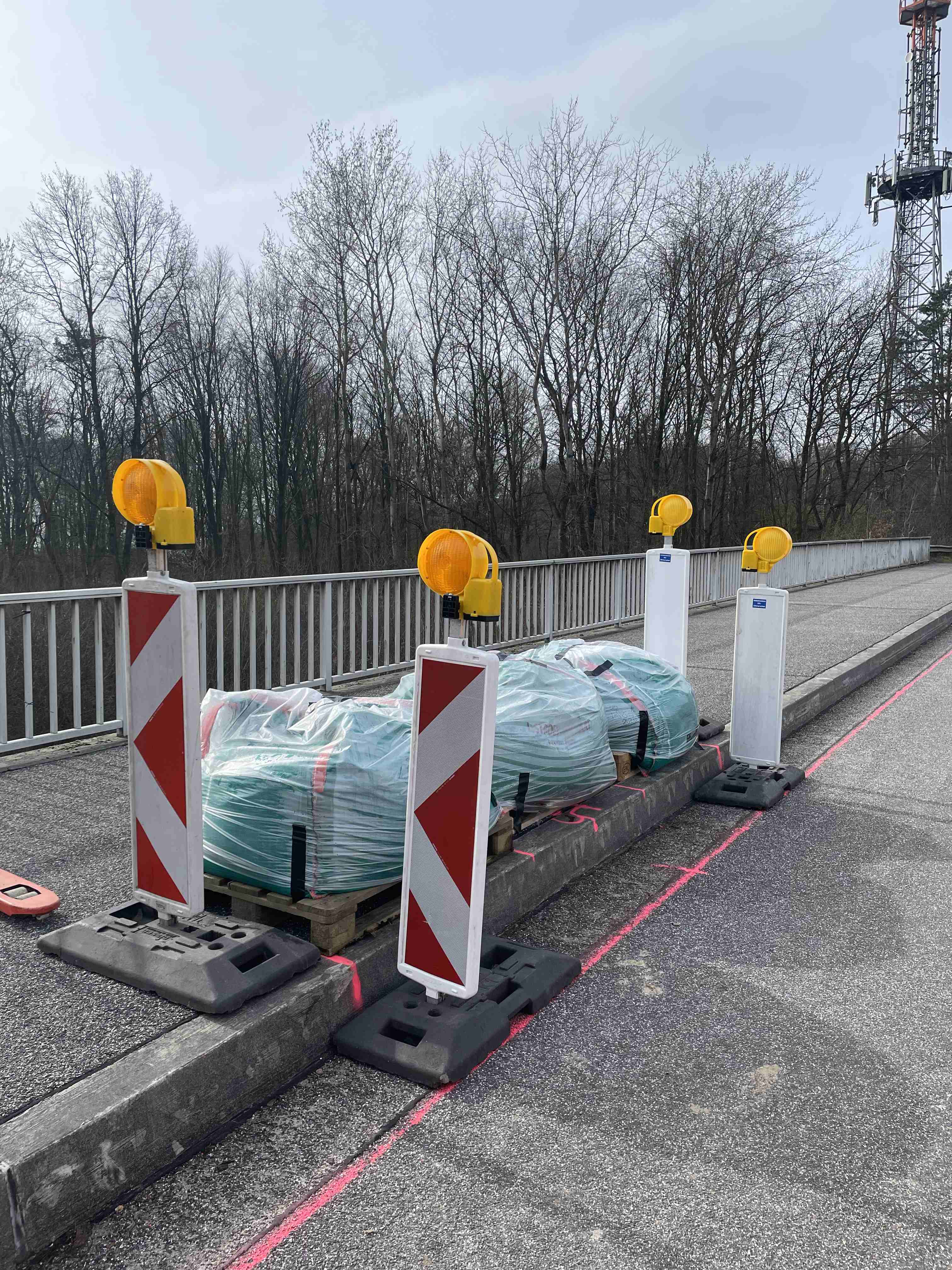}
    \caption{North view of concrete box girder bridge (left) and additional masses (middle and right).}
    \label{fig:North_view}
\end{figure}

The bridge is currently in operation, therefore, following consultation with the infrastructure operator, no information regarding the exact name and location of the bridge may be disclosed. The north view of the bridge is shown in Figure~\ref{fig:North_view} (left). It is a prestressed concrete bridge built with an open frame design and a box girder cross-section, constructed in 1972. The bridge has a main span of \SI{50}{m}; including the two transition structures of \SI{12.03}{m} and \SI{11.59}{m}, the instrumented longitudinal reference length is \SI{73.62}{m}. The bridge has a width of 10~m, accommodating a single lane for agricultural traffic and a pedestrian walkway. The bridge was initially designed for a traffic load corresponding to Bridge Class~30 according to the German standard DIN~1072 \citep{DIN_1072}, with a permissible axle load of \SI{10000}{kg}. Due to its location and usage, the bridge is expected to experience limited exposure to de-icing salts, which would lead to chloride-induced corrosion and freeze-thaw damage. For this reason, no further investigations or sensors for corrosion measurements were included in the monitoring system.  

The longitudinal structure comprises an open frame with cantilever arms on both sides and an arch system. 
The cross-section features a single-cell box girder with a maximum height of \SI{2.8}{m}, tapering in an arch-like manner to \SI{1.1}{m} at midspan. 
The box girder is accessible via two entry openings. Internal prestressing is installed in both longitudinal and transverse directions and grouted with mortar. Portland cement was predominantly used as a binder for the concrete. 

The bridge underwent major refurbishments in 1978 due to construction defects. 
Additional maintenance measures were carried out in 2010 and 2011, where a surface protection system (OS-C) was applied to the caps. 
According to inspection reports from 2018 and 2021, the bridge currently exhibits no or only minor defects, such as a \SI{0.01}{m} height offset at the expansion joint and concrete cracks in the caps and frame corners. 

%%%%%%%%%%%%%%%%%%%%%%%%%%%%%%%%%%%%%%%%%%%%%%%%
\section{Monitoring system}
%%%%%%%%%%%%%%%%%%%%%%%%%%%%%%%%%%%%%%%%%%%%%%%%

The SHM system is essential for evaluating the bridge's long-term performance and safety. 
Its primary goal is to continuously monitor structural behavior under various loads, detect early signs of damage, and provide near-real-time data for maintenance decisions and reliability assessments. 

\begin{figure}[!htb]
    \centering
    \includegraphics[width = .85\textwidth]{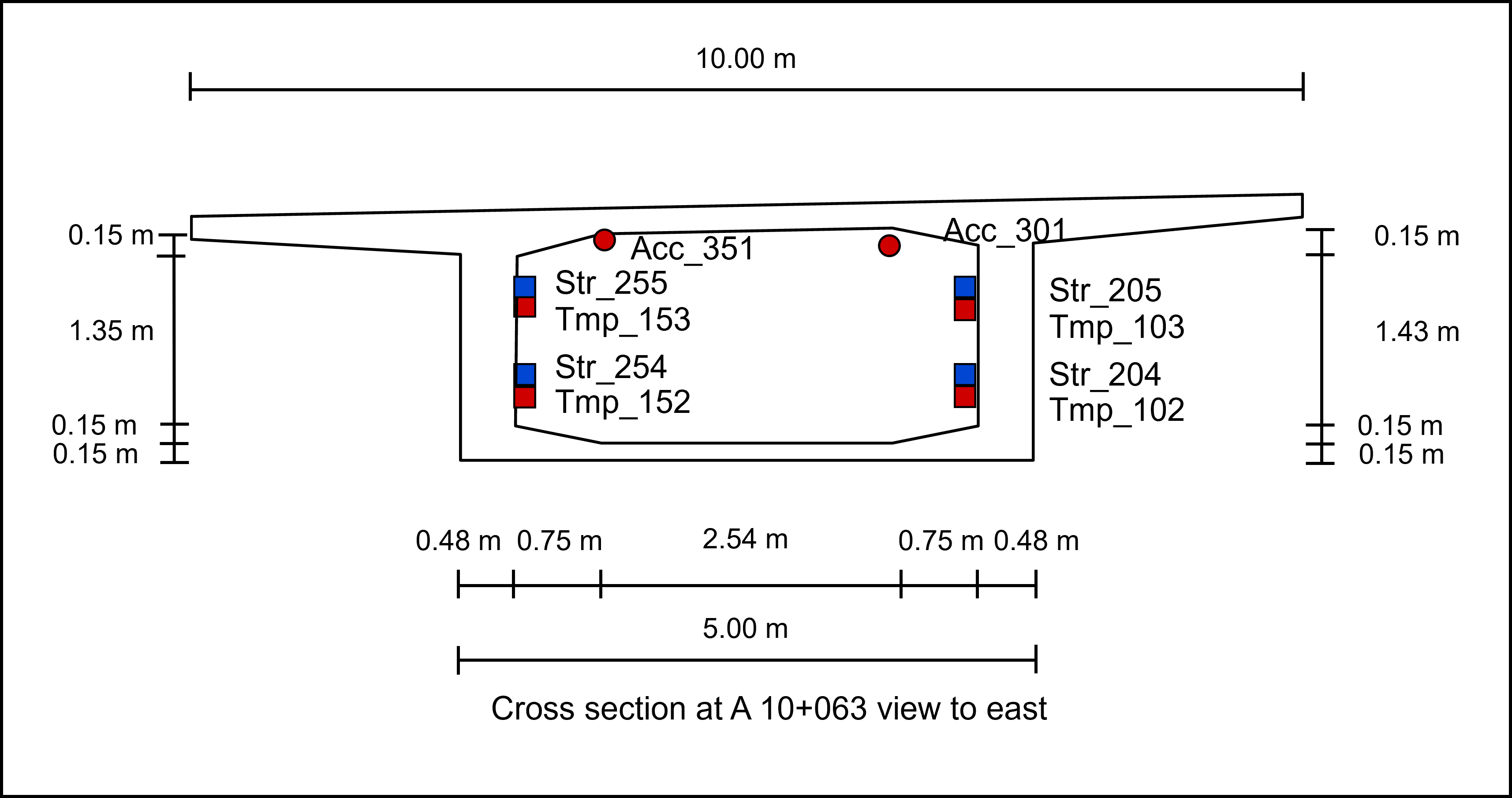}
    \caption{Cross-sectional instrumentation plan of the box girder bridge at A10+063.}
    \label{fig:Cross_Section}
\end{figure}

\begin{figure}[!htb]
    \centering
    \includegraphics[width=\linewidth]{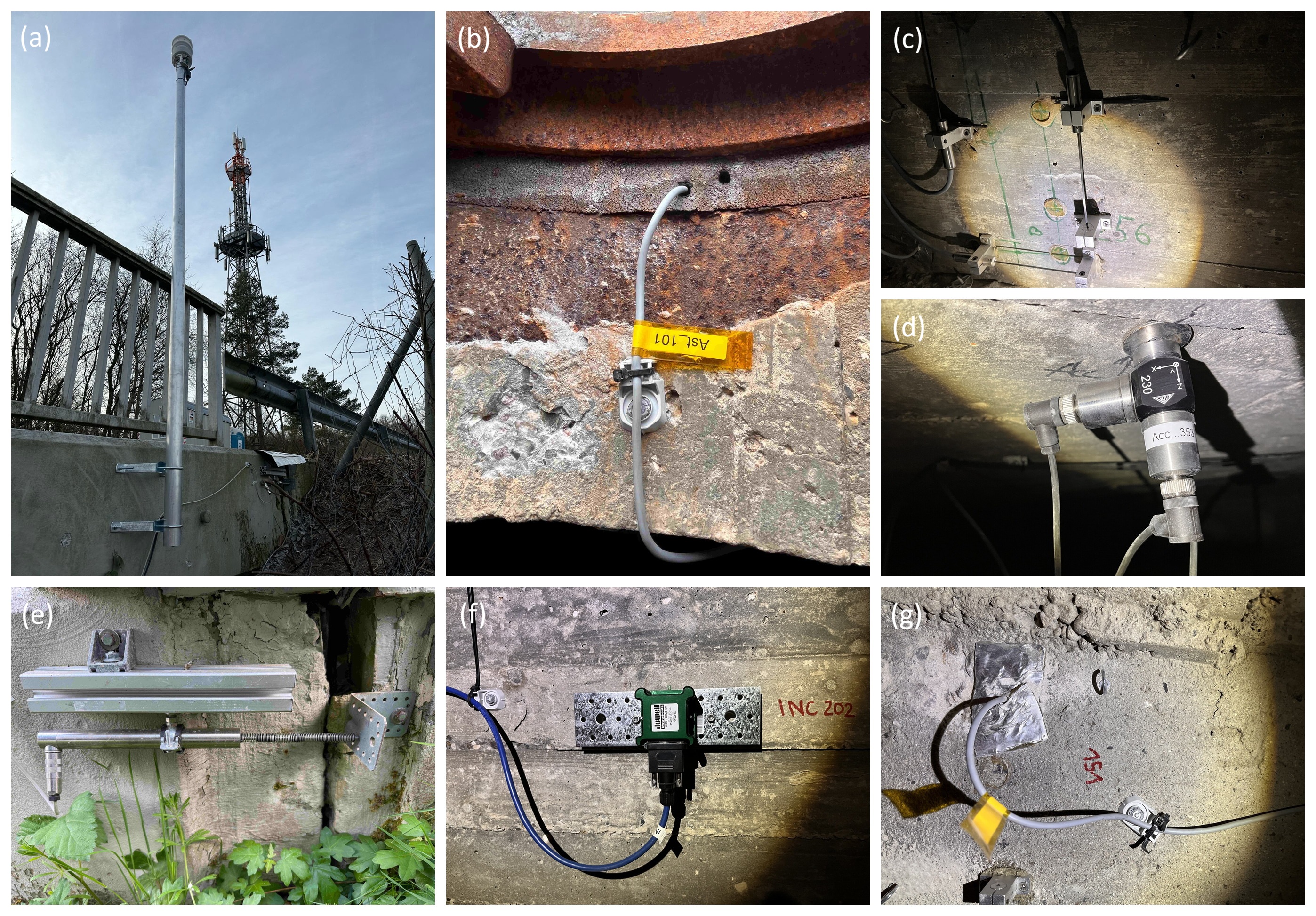}
    \caption{Mounting details of selected bridge-monitoring instrumentation, shown from left to right and top to bottom: 
    (a)~weather station WST~1, 
    (b)~asphalt-temperature probe Ast\_102, 
    (c)~strain sensors Str\_256Y and Str\_256Z, 
    (d)~accelerometer Acc\_353, 
    (e)~displacement transducer Dsp\_103, 
    (f)~inclinometer Inc\_202, and 
    (g)~temperature probe Tmp\_151.}
    \label{fig:Sensor_mounting_compilation}
\end{figure}

The monitoring system employs strain sensors, inclinometers, displacement transducers, accelerometers, temperature sensors, and weather stations. Sensor positions are given
in the top-view, side-view, and cross-sectional plans in Figures~\ref{fig:Top_View}--\ref{fig:Cross_Section}, and mounting details for each sensor type are shown in Figure~\ref{fig:Sensor_mounting_compilation}.
The drawings summarize the sensor positions to avoid overloading the drawings. 
For this reason, only one position is indicated in each case, even if there are actually multiple sensors, as is the case with the strain and acceleration sensors.
For precise localization, the spatial coordinates of the sensor positions were added to each sensor, which simplifies comparison with simulation results from finite element analyses. 
The coordinate origin is shown in Figure \ref{fig:Top_View}.
\begin{table}[!htb]
    \centering
    \caption{Overview of the long-term installed sensors, sampling rates, and mounting.}
    \label{tab:sensor_overview}
    \begin{threeparttable}
    \begin{tabular}{@{}l
        S[table-format=2.0]
        S[table-format=3.0]
        l@{}}
        \toprule
        \multicolumn{1}{c}{\textbf{Sensors}} &
        {\textbf{Qty. [pcs]}} &
        {\textbf{Rate [Hz]}} &
        \multicolumn{1}{c}{\textbf{Mounting}} \\
        \midrule
        Uniaxial accelerometers (Acc\_)       & 36 & 200 & Magnetic \\
        Inclinometer sensors (Inc\_)          & 8  & 100 & Screw clamp fastening \\
        Strain sensors (Str\_)                & 56 & 200 & Screw clamp fastening \\
        Displacement transducers (Dsp\_)      & 4  & 100 & Screw clamp fastening \\
        Weather stations (Wst\_)\tnote{*}     & 2  & 0.1  & Externally mounted \\
        Concrete temperature sensors (Tmp\_)  & 16 & 10  & Glued into the structure \\
        Asphalt temperature sensors (Ast\_)  & 2 & 10  & Glued into the structure \\
        Indoor air temperature sensors (IATmp\_)  & 2 & 10  & Free-hanging \\
        Humidity sensors (Hmd\_)  & 2 & 10  & Free-hanging \\
        \bottomrule
    \end{tabular}
    \begin{tablenotes}
        \footnotesize
        \item[*] Wind angle and speed, air temperature, relative humidity, and solar radiation.
    \end{tablenotes}
    \end{threeparttable}
\end{table}

\begin{minipage}{0.48\textwidth}
\begin{figure}[H]
    \centering
    \includegraphics[width = 2.6\textwidth, angle=90]{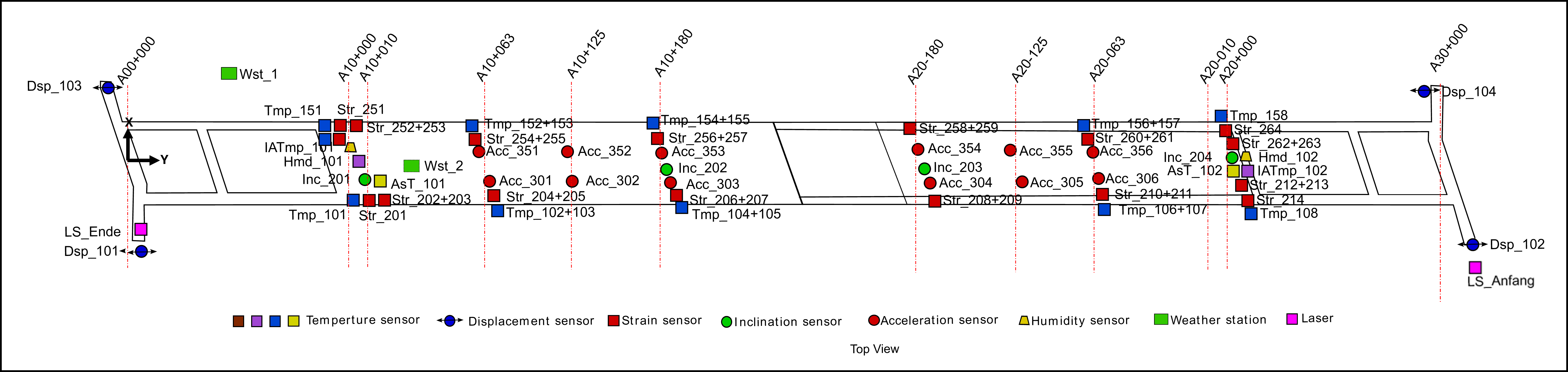}
    \caption{Top-view instrumentation plan of the box girder bridge.}
    \label{fig:Top_View}
\end{figure}
\end{minipage}
\hfill
\begin{minipage}{0.48\textwidth}
\begin{figure}[H]
    \centering
    \includegraphics[width = 2.6\textwidth, angle=90]{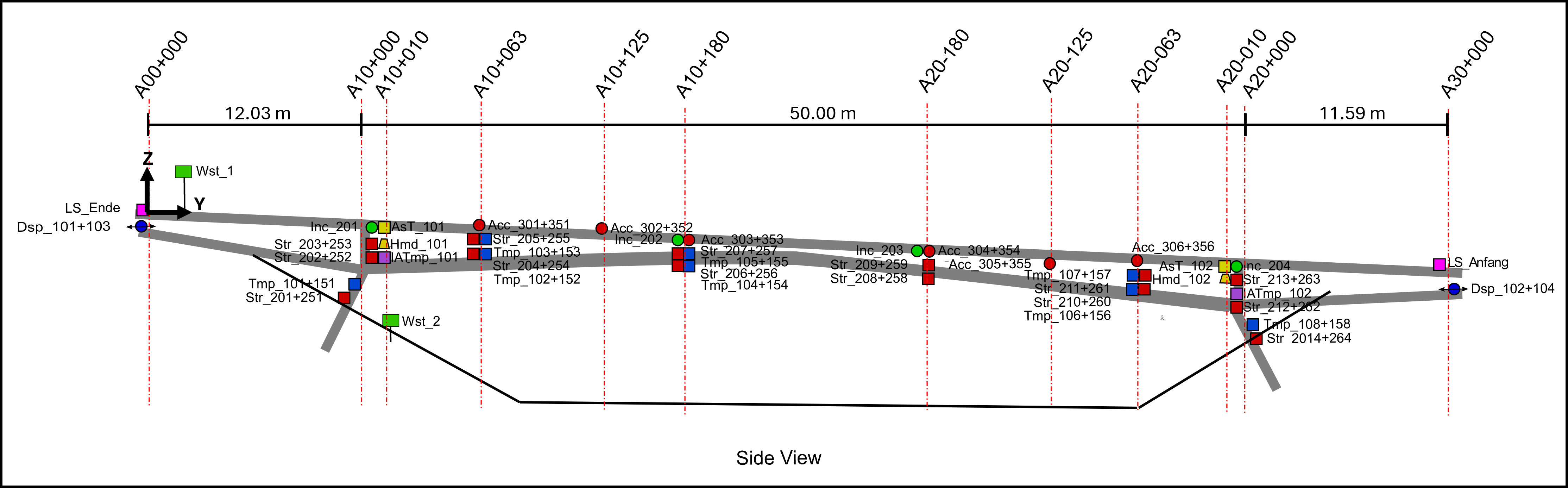}
    \caption{Side-view instrumentation plan of the box girder bridge.}
    \label{fig:Side_View}
\end{figure}
\end{minipage}

Strain sensors are strategically installed to measure the strain on concrete elements in three principal directions (x, y, z), distributed throughout the bridge. 
The strain measurement was performed using high-precision miniature inductive displacement transducers in a half-bridge configuration. 
For each measurement direction, a sensor with a corresponding signal amplifier was used and bolted to the structure. 
Due to a drain pipe in the north-east part of the bridge, it was not possible to mount the sensors on the wall to measure in the Z- and Y-directions. 
Instead, the sensors are placed on the underside of the concrete slab to measure the X- and Y-directions [Figure~\ref{fig:Sensor_mounting_compilation}(c)].
The coordinates and details of the sensors are shown in Table \ref{tab:long} (see appendix), \ref{tab:sensor_overview}, and \ref{tab:sensor_details}. 

\begin{table}[!htb]
    \centering
    \footnotesize
    \setlength{\tabcolsep}{2.3pt}
    \renewcommand{\arraystretch}{1.25}
    \caption{Measurement setup details and sensor specifications according to the manufacturers' data sheets.}
    \label{tab:sensor_details}
    \begin{tabularx}{\textwidth}{@{} A B C D E @{}}
        \toprule
        \textbf{Sensor type} &
        \textbf{Label} &
        \textbf{Manufacturer [Model]} &
        \textbf{Range} &
        \textbf{Specification} \\
        \midrule

        Acceleration, uniaxial &
        Acc\_3XX &
        Metra Mess- und Frequenztechnik [KS48C] &
        $\pm\SI{6}{g}$ &
        $\pm\SI{5}{\percent}$ \\

        Inclination, biaxial &
        Inc\_2XX &
        Althen Sensors \& Controls [JMI-200-D] &
        $\pm\SI{14.5}{\degree}$ &
        $\SI{0.002}{\degree}$ \\

        Weather station &
        Wst\_1XX &
        Lufft [WS700-UMB] &
        See data sheet &
        See data sheet \\

        Strain, uniaxial &
        Str\_2XX &
        Messotron [Wegaufnehmer WM] &
        $\pm\SI{0.5}{mm}$ &
        $\SI{80}{mV/V/mm}$ \\

        Displacement, uniaxial &
        Dsp\_1XX &
        Schreiber Messtechnik [SM40] &
        \SIrange{0}{100}{mm} &
        $\SI{0.25}{\percent}$ \\

        Temp., material &
        Tmp\_1XX &
        Thermokon [TF25-PT1000 T100 050.6] &
        \SIrange{-35}{100}{\degreeCelsius} &
        $\pm\SI{0.15}{\degreeCelsius}$ \\

        Temp., indoor air &
        IATmp\_1XX &
        Rotronic [D-M-HC2-Fühler-V1\_12] &
        \SIrange{-40}{60}{\degreeCelsius} &
        $\pm\SI{0.05}{\degreeCelsius}$ \\

        Temp., asphalt &
        Ast\_1XX &
        Thermokon [TF25-PT1000 T100 050.6] &
        \SIrange{-35}{100}{\degreeCelsius} &
        $\pm\SI{0.15}{\degreeCelsius}$ \\

        Humidity, air &
        Hmd\_1XX &
        Rotronic [D-M-HC2-Fühler-V1\_12] &
        \SIrange{0}{100}{\percent rH} &
        $\pm\SI{0.8}{\percent rH}$ \\

        Laser distance &
        LS\_XXX &
        Micro-Epsilon [optoNCDT ILR 2250] &
        \SIrange{0.05}{100}{m} &
        $\SI{0.1}{mm}$ \\

        \bottomrule
    \end{tabularx}
\end{table}
Inclinometers are positioned at support points and in the middle of the bridge to monitor angular changes or tilting, which can indicate alignment shifts or settlement issues (Figure~\ref{fig:Sensor_mounting_compilation}(f)).
Displacement transducers are installed at expansion joints to capture horizontal displacements, revealing abnormal movements that could compromise the bridge’s stability. In the process of configuring and preparing the sensors, the positive measurement direction for the plunger's inward movement was defined in such a manner that the expansion of the bridge is measured. 
The displacement sensors are screwed to an aluminum profile and pressed against a bracket screwed to the counterpart of the expansion joint [Figure~\ref{fig:Sensor_mounting_compilation}(e)].
The accelerometers are screwed on mounting cubes to enable them to be mounted securely and quickly. 
The cube is mounted on a steel plate system with magnets, which is dowelled beneath the concrete deck to measure vibration in three principal directions [Figure~\ref{fig:Sensor_mounting_compilation}(d)].

The flexibility of the steel plate system enables repositioning of the accelerometers as required for different research purposes, thereby enabling targeted dynamic assessments. Currently, the distance between the plates is 0.4 m, but this can be reduced by adding more plates.
Weather stations (WST) are also installed both above and below the bridge to monitor environmental parameters, including air temperature, humidity, wind speed, and direction. 
Temperature sensors embedded in the concrete elements, the asphalt, and the air of the box girder measure temperature fluctuations, provide insight into how thermal changes may affect bridge performance; see Table~\ref{tab:sensor_overview} and Figure \ref{fig:Sensor_mounting_compilation}. 
Thus, the installed monitoring system is well-suited to capturing valuable data during the load tests. 
The quantity of sensors was derived from the configuration of the bridge, with the objective of facilitating detailed monitoring with sufficient distances between measurement points. 
Almost all sensors were installed as planned, only the strain sensors in the northeast part of the bridge had to be adjusted due to the position of the drain pipes.
Environmental conditions are measured above (WST~1) and beneath (WST~2) the bridge. 
The data collected included global radiation, relative humidity (inside and outside of the bridge), air pressure, precipitation, and temperature (at bridge level, asphalt, and within the structure). 
The mounting of the weather station WST~1 is shown in Figure \ref{fig:Sensor_mounting_compilation}(a). 

The temperatures in different locations exposed to solar radiation provide a better understanding of the structure's temperature distribution and indicate parts with higher temperature-induced tension. 
Measuring temperature within the structure involved the use of glued-in Pt1000 temperature sensors with an installation length of \SI{0.05}{m}, bonded with Sikadur-31 CF Rapid adhesive mortar or HBK ABM 75 self-adhesive membrane. To mitigate the impact of radiation, the sensors were strategically mounted inside the box girder (Figure~\ref{fig:Sensor_mounting_compilation}(g)).

The sensors are named based on their measurement directions relative to the chosen global coordinate system: the Y-axis runs longitudinally along the bridge, the X-axis transversely, and the Z-axis vertically upward. The coordinate origin is located on the western side, centered at the bridge’s roadway crossing. 

All sensors were recorded using standardized IOLITE data acquisition units from DEWESOFT. 
Strain, inclination, and displacement were measured using IOLITE 8xLA modules, temperatures using IOLITE 8xRTD modules, and accelerations using KRYPTON 8xACC modules. The signals were consolidated in a single instance of Dewesoft X measurement software from Dewesoft, enabling time-synchronized measurement. 
Depending on the number and cross-section of the wires, the Lapp-Unitronic LIYCY product is used as a sensor cable. 
Only cables that are approved for the specific application are used. 
The cable jackets provide adequate UV protection and are designed to withstand mechanical stress (such as pulling cables over rough surfaces or edges). 
The cable lengths between the sensor and the data acquisition unit were limited to a maximum of \SI{25}{m} to minimize the impact of cable length on signal quality. 
To achieve this, a control cabinet was installed in each half of the box girder to consolidate the sensor signals, which were then transmitted to the measurement computer. The software allows the definition of sampling rates per measurement channel. The sampling rates are shown in Table~\ref{tab:sensor_overview}.

Three additional laser sensors were installed temporarily for the short-term static and dynamic load tests. 
While sensor locations are documented in the global coordinate system, selected load-test validation figures use a one-dimensional bridge station axis for truck positions. 
The station axis starts at A00+000 at the western transition and increases toward A30+000 at the eastern transition.
The position-tracking laser LS\_Front measured the truck's longitudinal position, which was used to verify the parking stations during the static test and to track the vehicle during the dynamic crossings. The other two sensors (LS\_Anfang and LS\_Ende) served as barriers at the bridge entrance and exit to record passage times and estimate velocity during the dynamic tests. 
The barrier timing is independent of the position-tracking laser; this redundancy allows the written test protocol to be verified and increases confidence in the recorded entry and exit times. All laser sensors were attached to the railing. See the appendix for further information.

%%%%%%%%%%%%%%%%%%%%%%%%%%%%%%%%%%%%%%%%%%%%%%%%
\section{Experimental procedure for load tests}\label{sec:experimental_procedure}
%%%%%%%%%%%%%%%%%%%%%%%%%%%%%%%%%%%%%%%%%%%%%%%%

%%%%%%%%%%%%%%%%%%%%%%%%%%%%%%%%%%%%%%%%%%%%%%%%
\subsection{Static Load Tests: Design and Execution}\label{subsec_static_load_tests}
%%%%%%%%%%%%%%%%%%%%%%%%%%%%%%%%%%%%%%%%%%%%%%%%

This section outlines the planning, preparation, and challenges of static loading tests. 
Two tests were conducted: a long-term test with Big Bags of sand placed on the bridge to simulate continuous stress, and a short-term test with a loaded truck parked on the bridge for the same purpose.

\begin{figure}[!htb]
   \includegraphics[width = .3\textwidth]{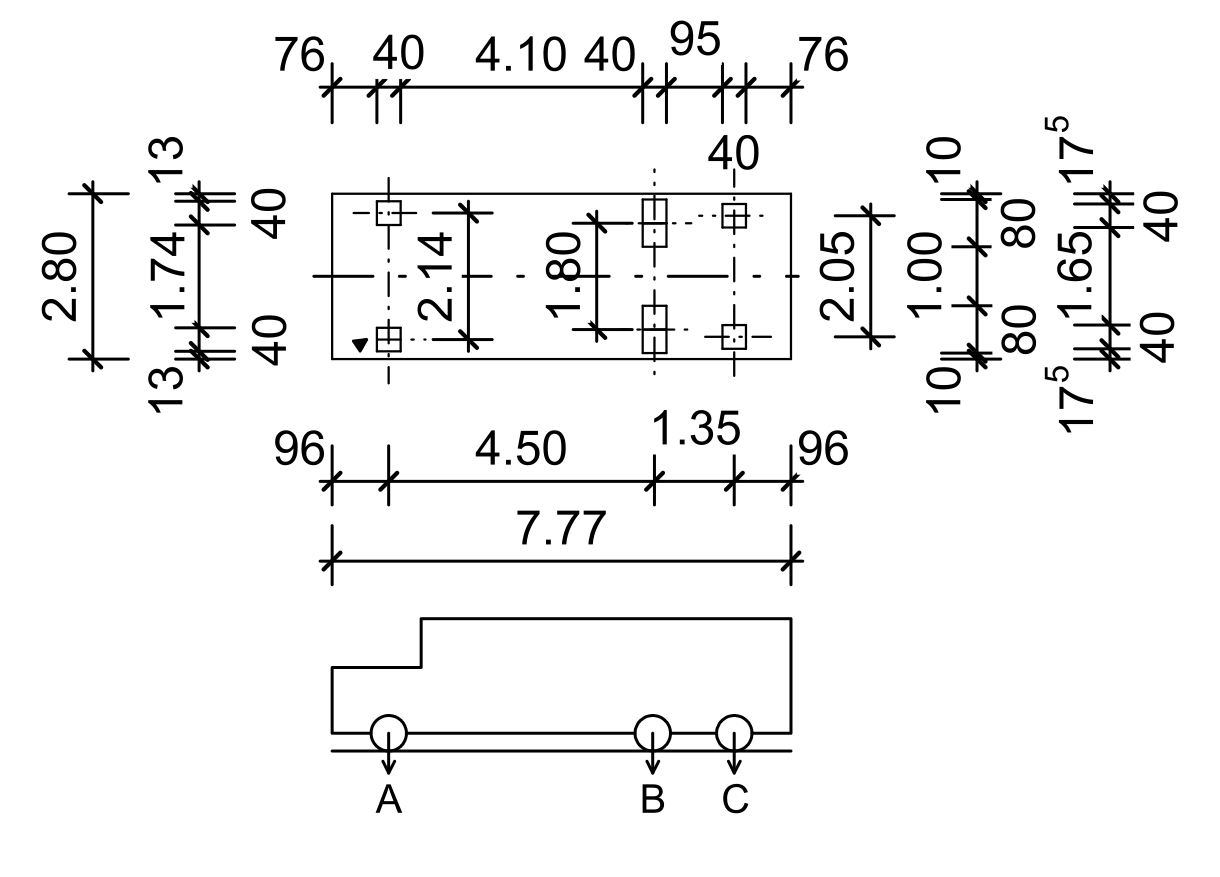}
    \includegraphics[width = .63\textwidth]{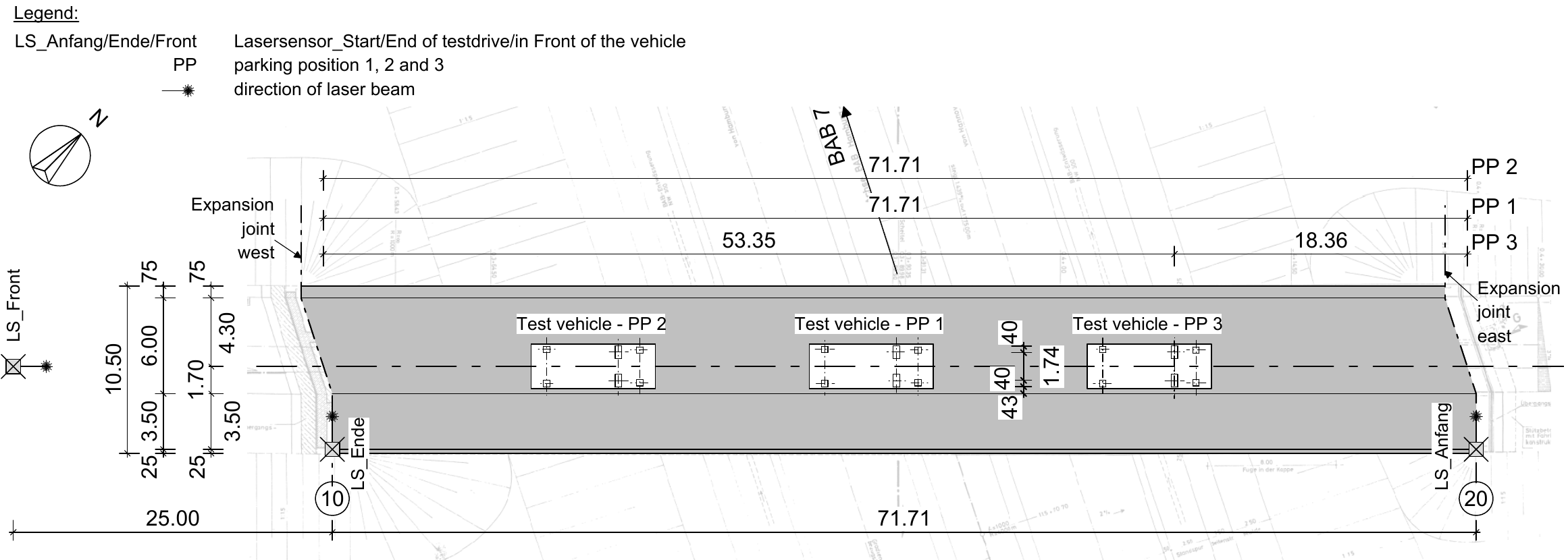}
    \caption{Axle dimensions [m] of the used truck (left). Overview of parking positions and laser sensors (right).}
    \label{fig:parking_positions}
\end{figure}

The truck, a three-axle truck (A, B, C axles from front to rear, Figure \ref{fig:parking_positions}~(left)), was weighed in two conditions: loaded and unloaded. When unloaded, it weighed \SI{12340}{kg}, and when loaded, the weight increased to \SI{21250}{kg}. 

In this station system, PP3 is closest to LS\_Anfang near the eastern side, PP1 is located near the bridge midpoint, and PP2 is closer to LS\_Ende and the western laser setup. 
Because station values increase from west to east, east-to-west truck passages appear as decreasing station values in the validation figures. 
The position-tracking laser LS\_Front was placed on the western side approximately \SI{25}{m} before A00+000 and measured the distance to the reflector mounted near the front centerline of the truck. 
Therefore, the front-reflector station was computed as the measured LS\_Front distance minus this \SI{25}{m} offset. 
The axle stations and truck-midpoint station were then obtained by adding the known truck-geometry offsets behind the front reflector. 
For the moving-passage response profiles, the plotted position is the estimated rear-axle station (Axle~3), obtained from the independent LS\_Anfang and LS\_Ende barrier crossings, the bridge station length, and the same truck geometry.

Table~\ref{tab:truck_loadings} in the appendix shows the axle loads, all within the permissible \SI{10000}{kg} per axle for the service load level. 
The short-term static load test was conducted with the loaded truck, and different tests were recorded in separate files for specific analysis. In contrast, the publication on Zenodo is in a consolidated HDF5 file. 
The file structure is described in the section titled "Data Format and Hierarchical Structure". 

\textit{Long-term static load test (extra masses).} 
To assess the monitoring system's sensitivity to small mass changes under varying environmental conditions, Big Bags filled with sand (approximately \SI{1600}{kg/m^3}) were placed on the bridge in three increasing steps: light, medium, and heavy at two positions (Table~\ref{tab:masses_overview}): first at the bridge midpoint on one side of the pedestrian path (Figure~\ref{fig:North_view}, middle and right), then at the quarter point on the western side (Figure~\ref{fig:parking_positions}, right). 
Moisture-related mass variations were minimized by enclosing the Big Bags.

\begin{table}[!htb]
    \centering
    \small
    \setlength{\tabcolsep}{1.8pt}
    \renewcommand{\arraystretch}{1.25}
    \caption{Overview of additional masses (in UTC).}
    \begin{tabular}{
        l
        c
        c
        S[table-format = 4.0]
        c
        c
    }
        \toprule
        \multicolumn{1}{c}{\textbf{Event}} &
        \textbf{Start} &
        \textbf{End} &
        {\textbf{Added weight [kg]}} &
        \textbf{Position (no.)} &
        \textbf{Days} \\
        \midrule
        Light weight  & 22.02.2024 09:58:00 & 03.03.2024 07:47:00 & 680  & Bridge midpoint (1) & 10 \\
        Medium weight & 03.03.2024 07:47:00 & 13.03.2024 09:33:00 & 1420 & Bridge midpoint (1) & 10 \\
        Heavy weight  & 13.03.2024 09:33:00 & 23.03.2024 09:48:00 & 2160 & Bridge midpoint (1) & 10 \\
        Light weight  & 23.03.2024 09:57:00 & 02.04.2024 07:41:00 & 680  & Quarter point (2)  & 10 \\
        Medium weight & 02.04.2024 07:41:00 & 12.04.2024 07:45:00 & 1420 & Quarter point (2)  & 10 \\
        Heavy weight  & 12.04.2024 07:45:00 & 25.04.2024 06:42:00 & 2160 & Quarter point (2)  & 14 \\
        \bottomrule
    \end{tabular}
    \label{tab:masses_overview}
\end{table}

\textit{Short-term static load test (parking positions).} 

\begin{figure}[!htb]
    \centering
    \includegraphics[width=0.49\linewidth]{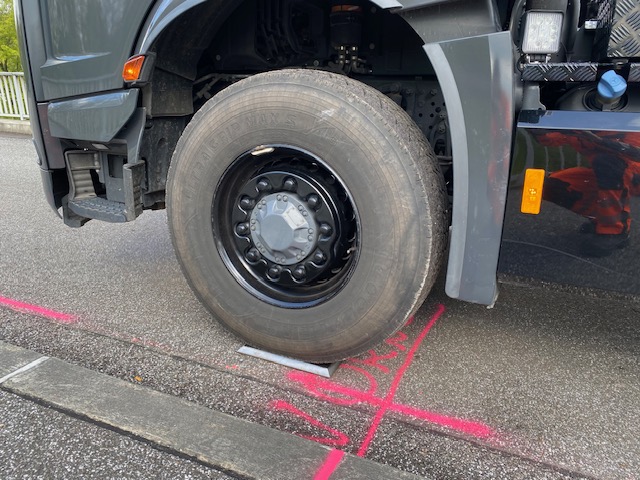}
    \includegraphics[width=0.49\linewidth]{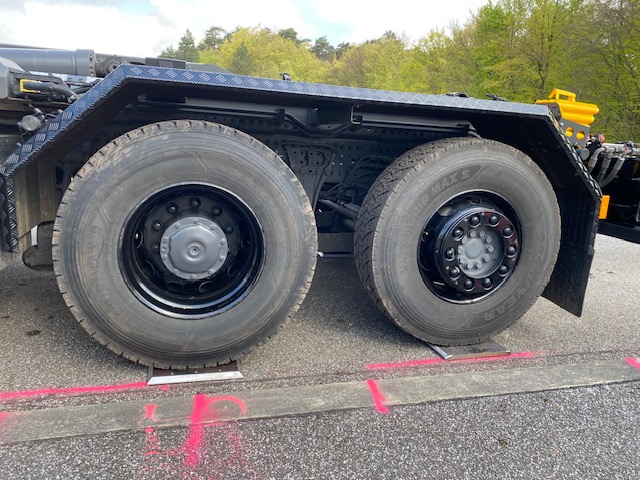} 
    \caption{Truck positioned on loading plates during static short-term load test.}
    \label{fig:file_parking_plates}
\end{figure}

The truck parking positions were used to simulate the design load case. 
The truck was driven to three positions at crawl speed to minimize dynamic effects, as recommended in the literature \citep{Lantsoght_2019, Alampalli.etal_2021, Schartner.etal_2022, Sanio.etal_2022}. 
The measurement in the first parking position was repeated three times for the design load-case simulation. 
At each position in Figure~\ref{fig:parking_positions}~(right) (PP1 in the center, PP2 and PP3 at the western and eastern quarter points, respectively), the truck parked on six steel load distribution plates (\SI{0.4}{m}~$\times$~\SI{0.4}{m}, \SI{0.025}{m} thick, total mass approximately \SI{187}{kg}) for at least ten minutes with the engine turned off (Figure~\ref{fig:file_parking_plates}). The driving lane was positioned near the center of the bridge to minimize torsional effects, and each drive was conducted from the eastern to the western end of the bridge, with both starting and ending points located off the bridge. Each loading position was repeated twice; parking times are listed in Table~\ref{tab:parkpos} (see appendix).

\begin{figure}[!htbp]
    \centering
    \includegraphics[width = 1\textwidth]{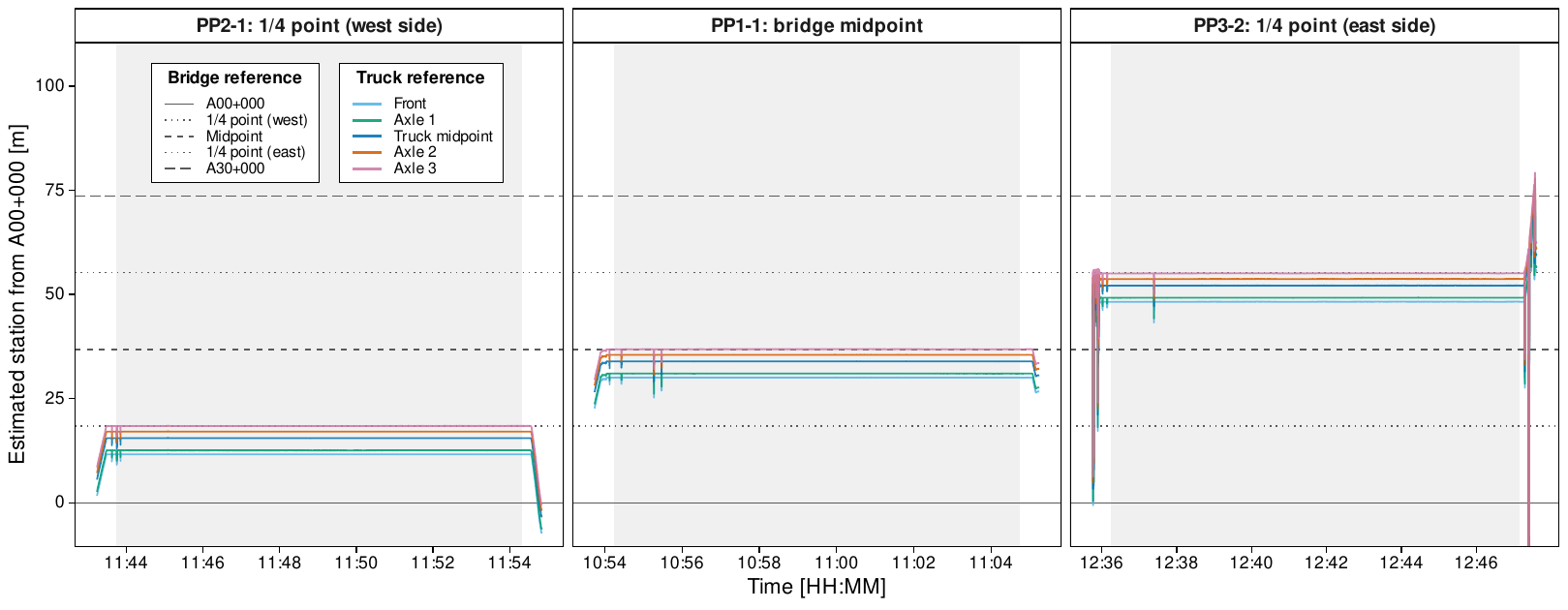}
    \caption{Laser-based validation of the short-term static parking positions. 
            Grey shading marks the measurement windows. 
            The estimated stations of the truck front, axles, and midpoint were computed from LS\_Front using the laser position on the western approach, approximately \SI{25}{m} outside A00+000, and the known truck geometry.
            }
    \label{fig:Laser_results}
\end{figure}

\begin{figure}[!htbp]
    \centering

    \begin{minipage}[t]{0.785\textwidth}
        \vspace{0pt}
        \centering
        \includegraphics[width=\linewidth]{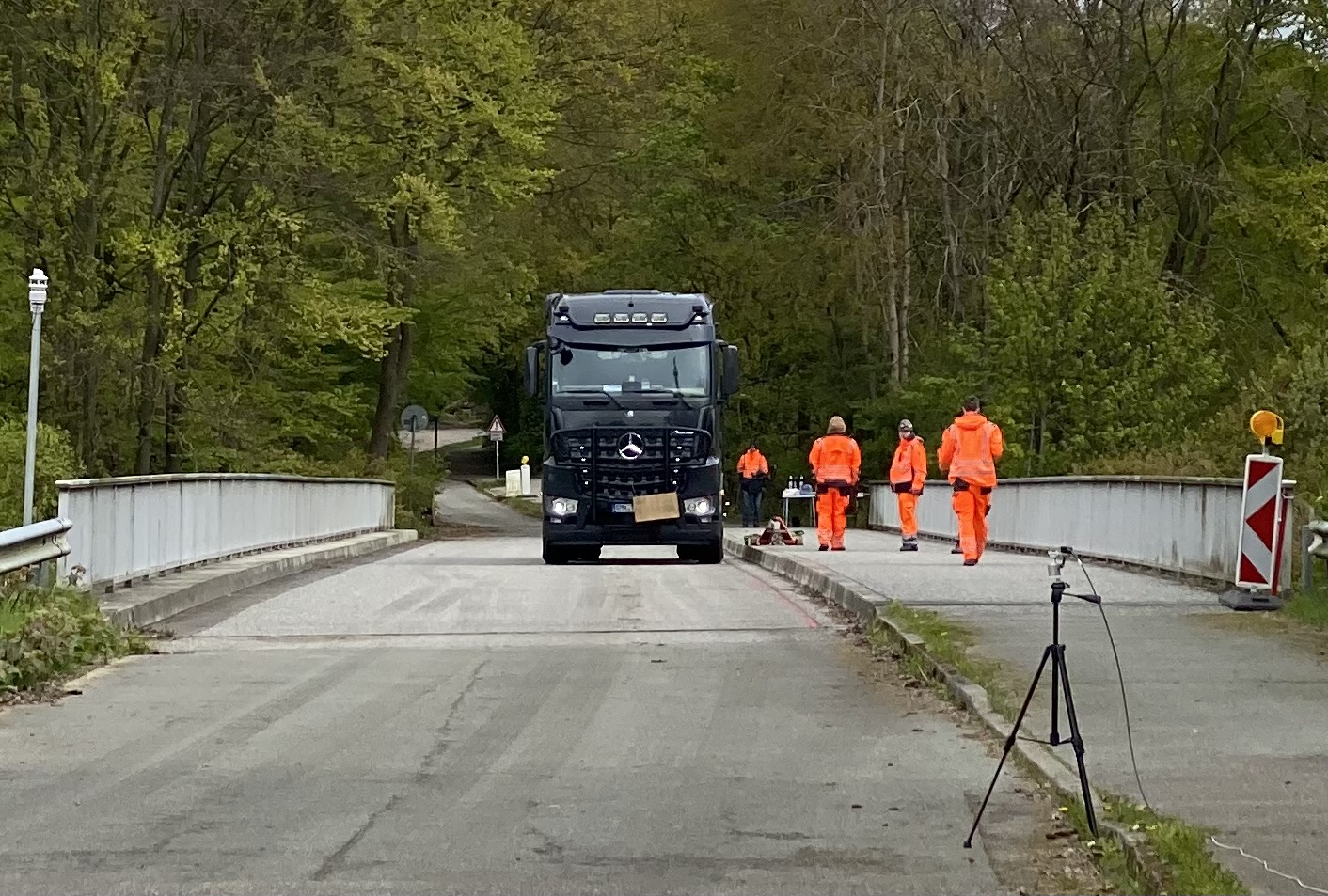}
    \end{minipage}
    \hspace{0.005\textwidth}
    \begin{minipage}[t]{0.2\textwidth}
        \vspace{0pt}
        \centering
        \includegraphics[angle=270,width=0.97\linewidth]{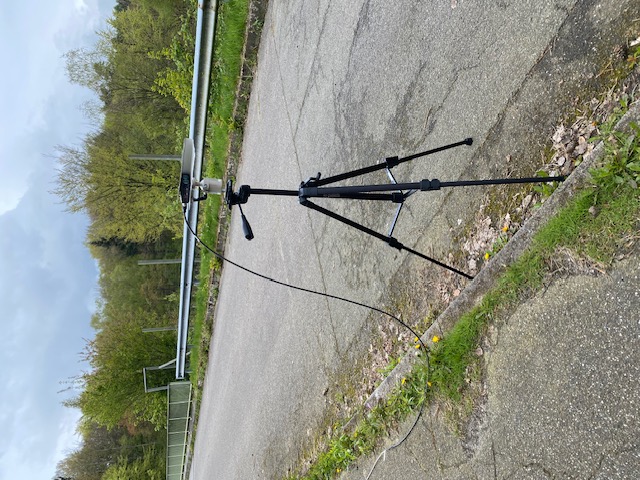}

        \vspace{2mm}

        \includegraphics[angle=270,width=0.97\linewidth]{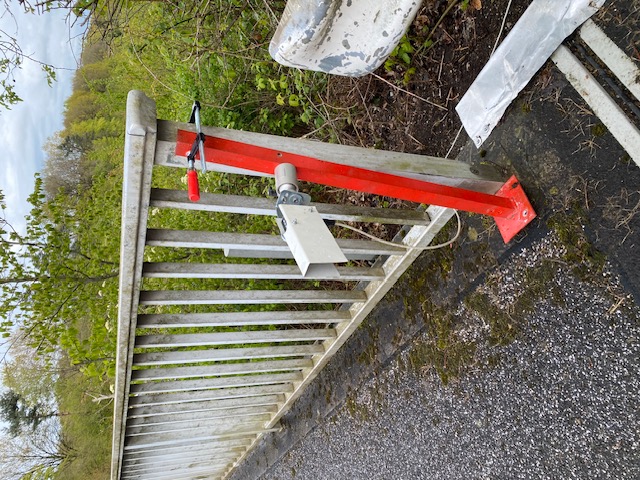}
    \end{minipage}

    \caption{
            Field setup for the laser-based measurement system during the short-term static and dynamic bridge load tests. 
            The overview image (left) shows the test vehicle on the bridge with a reflective target mounted at the front of the truck, enabling estimation of the truck position during the crossing; the detail images (right) show the tripod-mounted laser unit and the laser barrier for vehicle passage-time measurement.
            }
    \label{fig:file_laser_setup}
\end{figure}

For validation, the front laser sensor LS\_Front was used to identify the start and end of the parking intervals and to estimate the stations of the truck reference points (Figure~\ref{fig:Laser_results}). The stable plateaus confirm the parking intervals and show that the middle axle (Axle~2) is aligned with PP2, PP1, and PP3 with a small offset. 
Short transients at the window edges indicate vehicle positioning and departure phases and were not used to define the parked station.
The laser measurement setup used to record truck position and crossing times is shown in Figure~\ref{fig:file_laser_setup}.

The laser-based station estimate has an absolute uncertainty of the order of \SI{1}{m}, dominated by the choice of truck reference point (the axles are \SI{1.35}{m} apart) and the LS\_Front datum (the \SI{25}{m} offset before A00+000 and the laser-barrier placement). 
This affects only the absolute along-bridge position; the response magnitudes, repeatability, and neutral-axis estimates are unaffected, as they do not depend on the absolute station.

\FloatBarrier

%%%%%%%%%%%%%%%%%%%%%%%%%%%%%%%%%%%%%%%%%%%%%%%%
\subsection{Dynamic Load Test: Design and Execution}
%%%%%%%%%%%%%%%%%%%%%%%%%%%%%%%%%%%%%%%%%%%%%%%%

Dynamic load tests were conducted by driving the same three-axle truck across the bridge in loaded and unloaded configurations at four speeds based on the literature \citep{Sanio.etal_2022, Schartner.etal_2022, Lantsoght_2019, Alampalli.etal_2021}. 
The lowest speed (\SI{1.4}{m/s}) approximates quasi-static conditions, and \SI{13.9}{m/s} is the upper bound imposed by the posted speed limit. 
The intermediate speeds of \SI{5.6}{m/s} and \SI{11.1}{m/s} were selected following \citet{Sanio.etal_2022} and \citet{Schartner.etal_2022} to ensure variability in signal-to-noise ratio and to assess non-linearity in the structural response. 
In the loaded condition, crossings were performed at all four speeds; in the unloaded condition, only the two lowest speeds were used. 
The crawl-speed tests were repeated ten times to reduce uncertainty in comparisons; all other speed--weight combinations were repeated five times.
During the crawl-speed crossings, the driver followed a person walking along the pedestrian walkway, which set a steady walking-pace velocity (\SI{1.4}{m/s}) and minimized speed variation during the crossing.
The rear-axle station was obtained by fitting a constant-speed (linear) time--position model to the laser-barrier axle-crossing anchors, using the fixed truck axle geometry; the fit residual quantifies the position uncertainty.
The steady walking pace of the crawl-speed crossings makes this constant-speed assumption well justified.
The time periods of all measurement files are listed in Table~\ref{tab:passings}; the third crawl-speed test is incomplete due to a late measurement start, but the partial data are included in the dataset.

\begin{table}[!htb]
    \centering
    \footnotesize
    \setlength{\tabcolsep}{3.0pt}
    \renewcommand{\arraystretch}{1.18}
    \caption{Overview of measurement-file time periods for the vehicle passages (UTC).}
    \label{tab:passings}

    \begin{tabular*}{\linewidth}{@{\extracolsep{\fill}} l T T l N
                                      S[table-format=2.1] T T l N @{}}
    \toprule
    \multicolumn{1}{c}{\textbf{Event}} &
    \multicolumn{1}{c}{\textbf{Start}} &
    \multicolumn{1}{c}{\textbf{End}} &
    \multicolumn{1}{c}{\textbf{Case}} &
    \multicolumn{1}{c}{\textbf{No.}} &
    \multicolumn{1}{c}{\textbf{Event} } &
    \multicolumn{1}{c}{\textbf{Start}} &
    \multicolumn{1}{c}{\textbf{End}} &
    \multicolumn{1}{c}{\textbf{Case}} &
    \multicolumn{1}{c}{\textbf{No.}} \\
    \midrule

    Crawling & 08:31.45 & 08:33.15 & unloaded & 1  & 5.6 \si{\metre\per\second}  & 09:48.35 & 09:49.15 & loaded   & 1 \\
    Crawling & 08:34.59 & 08:36.13 & unloaded & 2  & 5.6 \si{\metre\per\second}  & 09:58.07 & 09:58.55 & loaded   & 2 \\
    Crawling & 08:37.24 & 08:38.32 & unloaded & 3  & 5.6 \si{\metre\per\second}  & 10:00.14 & 10:01.53 & loaded   & 3 \\
    Crawling & 08:39.35 & 08:40.48 & unloaded & 4  & 5.6 \si{\metre\per\second}  & 10:02.52 & 10:03.38 & loaded   & 4 \\
    Crawling & 08:42.29 & 08:43.40 & unloaded & 5  & 5.6 \si{\metre\per\second}  & 10:04.43 & 10:05.31 & loaded   & 5 \\
    Crawling & 08:45.15 & 08:46.29 & unloaded & 6  & 11.1 \si{\metre\per\second} & 10:06.54 & 10:07.31 & loaded   & 1 \\
    Crawling & 08:48.19 & 08:49.38 & unloaded & 7  & 11.1 \si{\metre\per\second} & 10:08.45 & 10:09.30 & loaded   & 2 \\
    Crawling & 08:50.31 & 08:51.53 & unloaded & 8  & 11.1 \si{\metre\per\second} & 10:10.30 & 10:11.06 & loaded   & 3 \\
    Crawling & 08:53.03 & 08:54.25 & unloaded & 9  & 11.1 \si{\metre\per\second} & 10:12.02 & 10:12.45 & loaded   & 4 \\
    Crawling & 08:55.18 & 08:56.33 & unloaded & 10 & 11.1 \si{\metre\per\second} & 10:13.37 & 10:14.24 & loaded   & 5 \\
    Crawling & 08:57.27 & 08:58.49 & unloaded & 11 & 13.9 \si{\metre\per\second} & 10:19.57 & 10:20.46 & loaded   & 1 \\

    \addlinespace[0.25em]

    Crawling & 09:27.05 & 09:28.28 & loaded   & 1  & 13.9 \si{\metre\per\second} & 10:22.34 & 10:23.21 & loaded   & 2 \\
    Crawling & 09:29.23 & 09:30.40 & loaded   & 2  & 13.9 \si{\metre\per\second} & 10:24.52 & 10:25.33 & loaded   & 3 \\
    Crawling & 09:31.24 & 09:32.38 & loaded   & 3  & 13.9 \si{\metre\per\second} & 10:26.53 & 10:27.34 & loaded   & 4 \\
    Crawling & 09:33.25 & 09:34.40 & loaded   & 4  & 13.9 \si{\metre\per\second} & 10:28.51 & 10:29.33 & loaded   & 5 \\
    Crawling & 09:35.18 & 09:36.35 & loaded   & 5  & 5.6 \si{\metre\per\second}  & 13:17.13 & 13:17.27 & unloaded & 1 \\
    Crawling & 09:37.17 & 09:38.33 & loaded   & 6  & 5.6 \si{\metre\per\second}  & 13:19.04 & 13:19.19 & unloaded & 2 \\
    Crawling & 09:39.13 & 09:40.36 & loaded   & 7  & 5.6 \si{\metre\per\second}  & 13:20.31 & 13:20.45 & unloaded & 3 \\
    Crawling & 09:41.18 & 09:42.33 & loaded   & 8  & 5.6 \si{\metre\per\second}  & 13:21.51 & 13:22.05 & unloaded & 4 \\
    Crawling & 09:43.18 & 09:44.35 & loaded   & 9  & 5.6 \si{\metre\per\second}  & 13:23.12 & 13:23.26 & unloaded & 5 \\
    Crawling & 09:45.01 & 09:46.25 & loaded   & 10 &
    \multicolumn{5}{c@{}}{} \\

    \bottomrule
    \end{tabular*}
\end{table}

\FloatBarrier

%%%%%%%%%%%%%%%%%%%%%%%%%%%%%%%%%%%%%%%%%%%%%%%%
\section{Load-Test Data Validation and Quality Assessment}
%%%%%%%%%%%%%%%%%%%%%%%%%%%%%%%%%%%%%%%%%%%%%%%%

This section documents validation and data-quality checks for the main measured response quantities. 
The checks are intended to verify time alignment, sensor response consistency, repeatability, and agreement with independently measured vehicle-position information. 
They do not replace a full quantitative model-validation study against a calibrated finite element model; rather, they provide transparent evidence that the published data are internally consistent and physically interpretable. 
The evaluation includes acceleration and strain measurements, as well as horizontal displacements at expansion joints. 
The inclinometer channels are included in the data set but are not shown in a separate response figure because the selected load cases were not expected nor intended to produce measurable rotations above the relevant noise level.

\begin{figure}[!htbp]
    \centering
    \includegraphics[width=\textwidth]{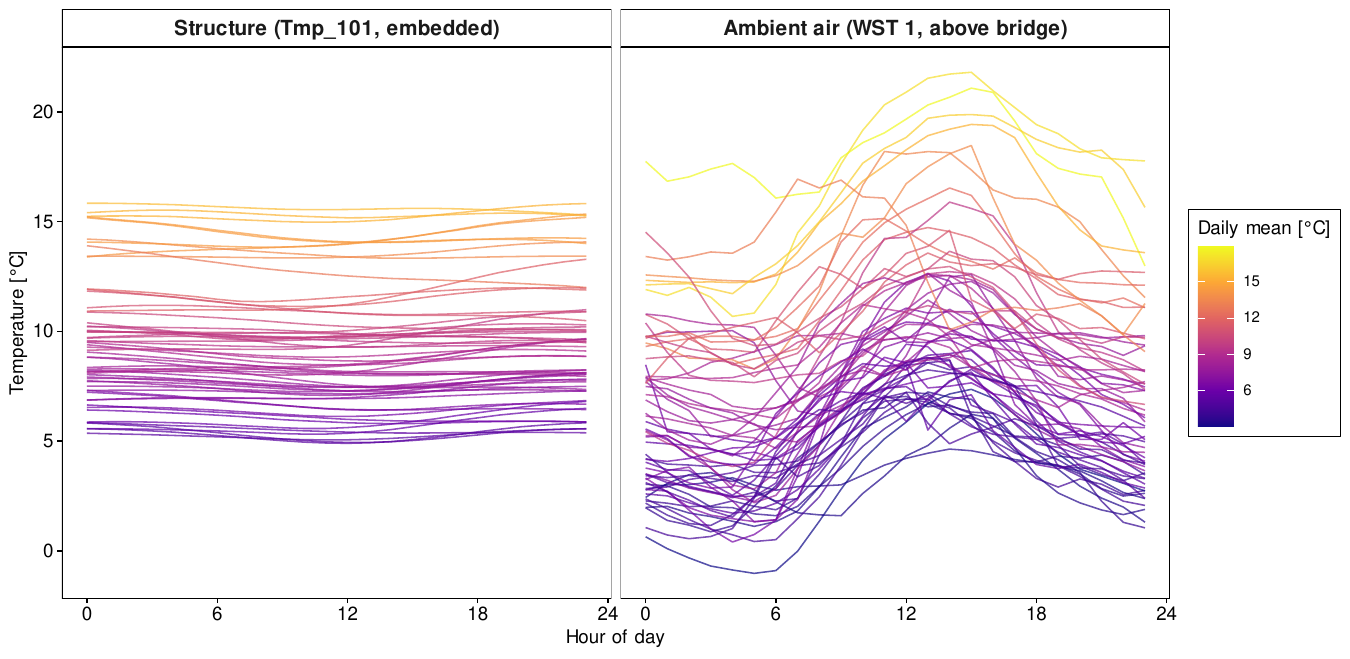}
    \caption{Hourly temperature profiles for each day of the long-term static load test campaign, one line per day coloured by that day's mean temperature. Left: embedded structure temperature (Tmp\_101, at a support); right: ambient air temperature (weather station WST~1, above the bridge).}
    \label{fig:temperature_profiles}
\end{figure}

The thermal environment spanned during the long-term static load test campaign is summarized in Figure~\ref{fig:temperature_profiles}, which shows the hourly temperature profiles for each day, separated into the embedded structure temperature (Tmp\_101) and the ambient air temperature measured by weather station WST~1 above the bridge. 
Over the campaign, the daily mean temperature ranged from approximately \SI{5}{\celsius} to \SI{16}{\celsius} for the embedded structure and from \SI{3}{\celsius} to \SI{18}{\celsius} for the ambient air. 
The median daily range of the hourly-mean temperature was only \SI{0.8}{\celsius} for the embedded structure, compared with \SI{6.4}{\celsius} for the ambient air; that is, the thermal mass of the structure damps the daily cycle by roughly a factor of eight. 
This damping, together with the associated thermal lag, is the physical origin of a within-day hysteresis between the structural response and temperature: because the structure lags the air temperature, an instantaneous temperature reading is not a one-to-one predictor of the structural response, which is relevant when the data are used for environmental-effect compensation and confounder-aware monitoring \citep{Wittenberg.etal_2025, Neumann.etal_2026a}.

%%%%%%%%%%%%%%%%%%%%%%%%%%%%%%%%%%%%%%%%%%%%%%%%
\subsection{Acceleration and modal response}
%%%%%%%%%%%%%%%%%%%%%%%%%%%%%%%%%%%%%%%%%%%%%%%%

\begin{figure}[!htbp]
    \centering
    \includegraphics[width=.8\textwidth]{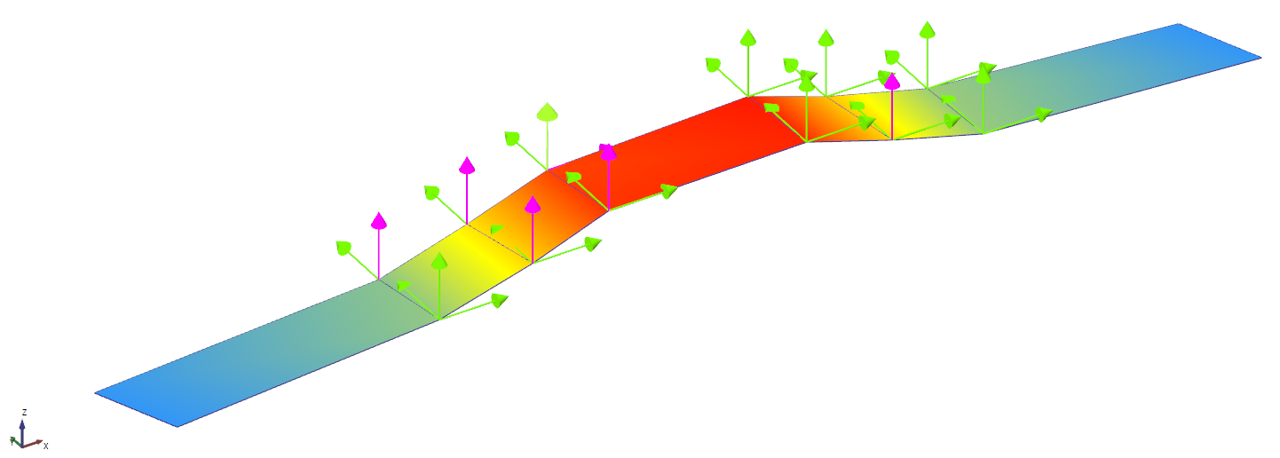}
    \includegraphics[width=.8\textwidth]{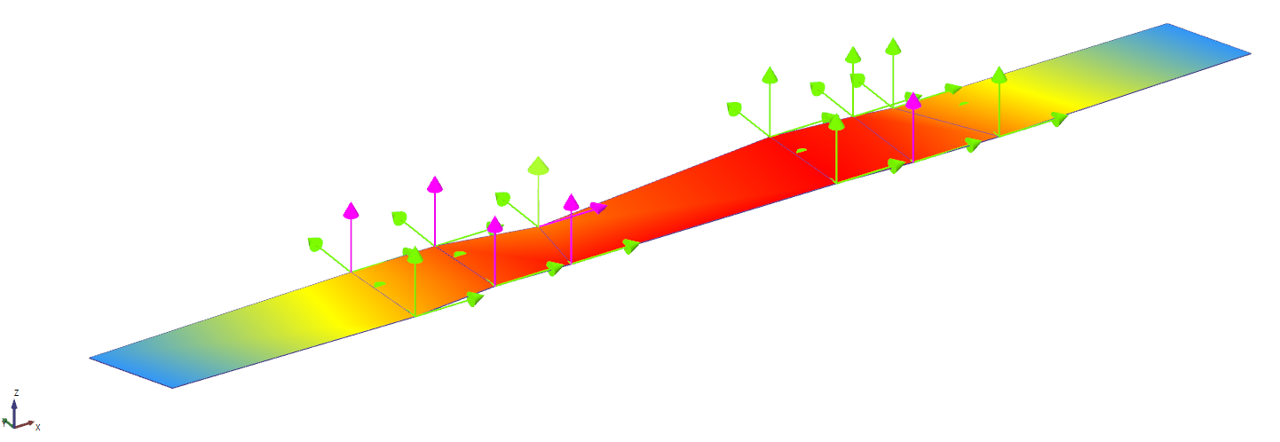} 
    \includegraphics[width=.8\textwidth]{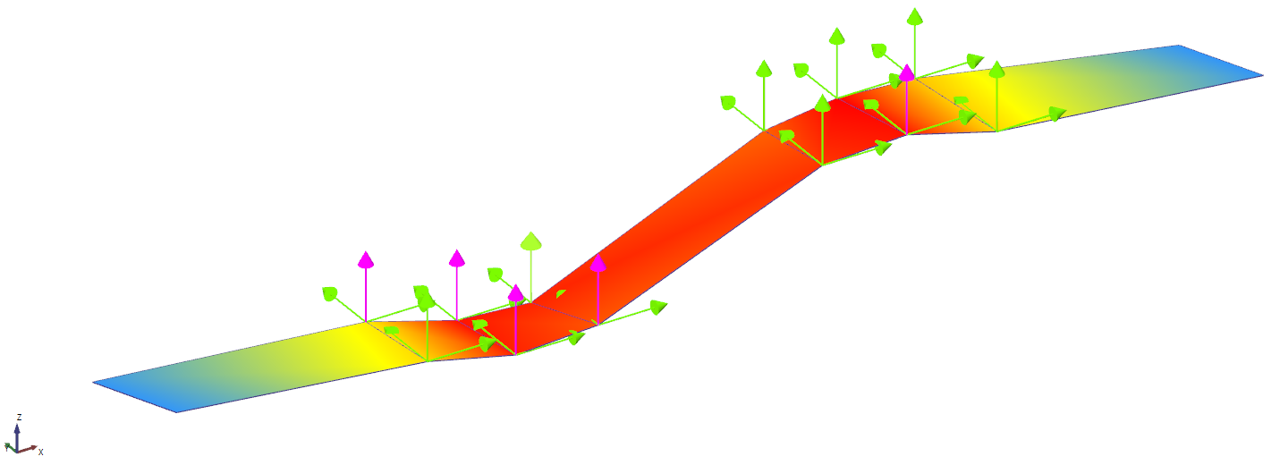} 
    \caption{
        Modal parameters at the reference state identified using Operational Modal Analysis (OMA): global bending mode ($f=\SI{3.06}{\hertz}$, $\zeta=\SI{2.02}{\percent}$; top), torsional mode ($f=\SI{3.95}{\hertz}$, $\zeta=\SI{3.5}{\percent}$; middle), and bending mode ($f=\SI{6.38}{\hertz}$, $\zeta=\SI{3.6}{\percent}$; bottom).}
    \label{fig:reference_states}
\end{figure}

Acceleration data were assessed using three complementary checks: (i) reference-state modal identification, (ii) frequency-domain inspection during dynamic truck passages, and (iii) long-term tracking of modal parameters during the added-mass test. For this purpose, Operational Modal Analysis (OMA) was performed using \SI{300}{s}  vibration data segments extracted at \SI{1800}{s} intervals. The analysis was conducted using the ARTeMIS Modal Pro 7.2 software package \citep{ARTeMIS_2022}, which estimates natural frequencies ($f$), damping ratios ($\zeta$), and mode shapes from ambient vibration measurements.
Figure \ref{fig:reference_states} depicts the reference state's global bending and torsional modes from January 1 to February 22, 2024. 
The results identified the first bending mode at \SI{3.06}{Hz} and the first torsional mode at \SI{3.95}{Hz}.

\begin{figure}[!htbp]
    \centering
    \includegraphics[width = 1\textwidth]{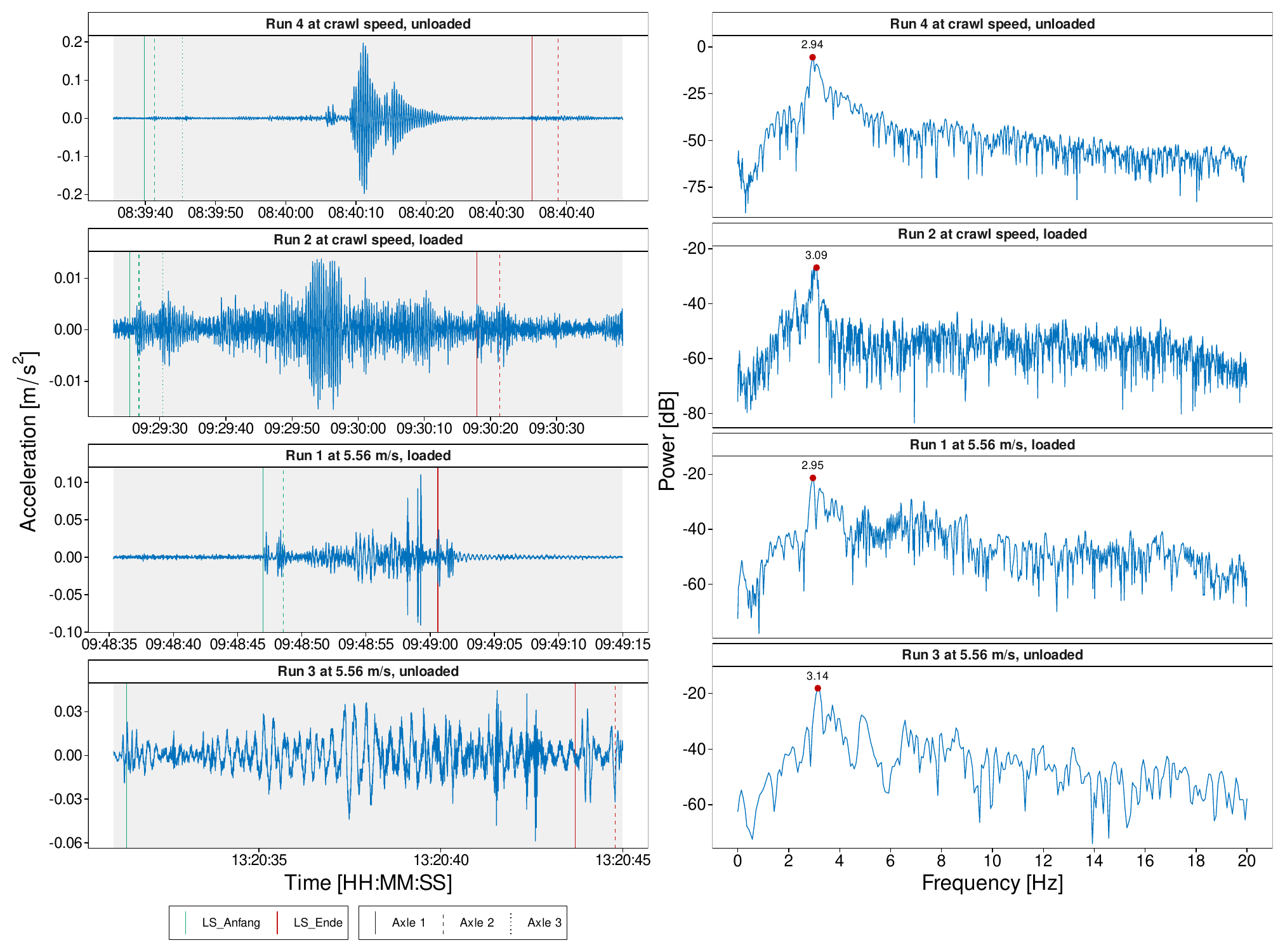}
    \caption{
            Representative vibration signals measured by sensor Acc\_353 in the $Z$-direction and corresponding frequency spectra for different truck speeds and loading conditions. 
            Grey shading marks the tabulated vehicle-passing windows. 
            Green and red vertical lines indicate LS\_Anfang and LS\_Ende axle-crossing candidates, respectively; line types distinguish the detected axle candidates.
            }
    \label{fig:typical_vibration_signal}
\end{figure}

As part of the dataset validation, plausibility checks on the vibration data were performed using time-domain inspection and Fast Fourier Transform (FFT) analysis. 
The vibration signals were checked for anomalies such as spikes, dropouts, clipping, drift, and unexpected signal morphology. 
Figure~\ref{fig:typical_vibration_signal} shows representative vibration signals measured by sensor Acc\_353 in the $Z$-direction at the middle of the bridge, together with the corresponding FFT spectra for different truck speeds and loading conditions. 
The grey shaded areas mark the tabulated vehicle-passing windows, while the vertical green and red lines indicate axle-crossing candidates detected by LS\_Anfang and LS\_Ende, respectively. 
This provides a consistency check between the passing-time metadata, the laser-based axle detections, and the measured acceleration response. 
The spectra show dominant peaks around the first resonance frequency, indicating that the dynamic tests successfully excited the first natural frequency. 
This supports further dynamic evaluations based on natural-frequency estimates derived from the generated data.

The acceleration checks, therefore, combine two complementary pieces of modal evidence. 
First, OMA of the reference-state monitoring data identified stable modal frequencies for the first bending and torsional modes (Figure~\ref{fig:reference_states}). 
And second, the FFT spectra from the dynamic truck passages show a dominant peak near the first identified eigenfrequency, indicating that the moving-load tests excited the first bridge mode (Figure~\ref{fig:typical_vibration_signal}). 
Together, these results provide a modal consistency check between the monitoring data, the operational modal analysis, and the dynamic load-test response. 
This comparison is used as a plausibility check for the acceleration data and modal-tracking results, not as a full calibration or validation of the finite element model.

\begin{figure}[!htbp]
    \centering
    \includegraphics[width = \textwidth]{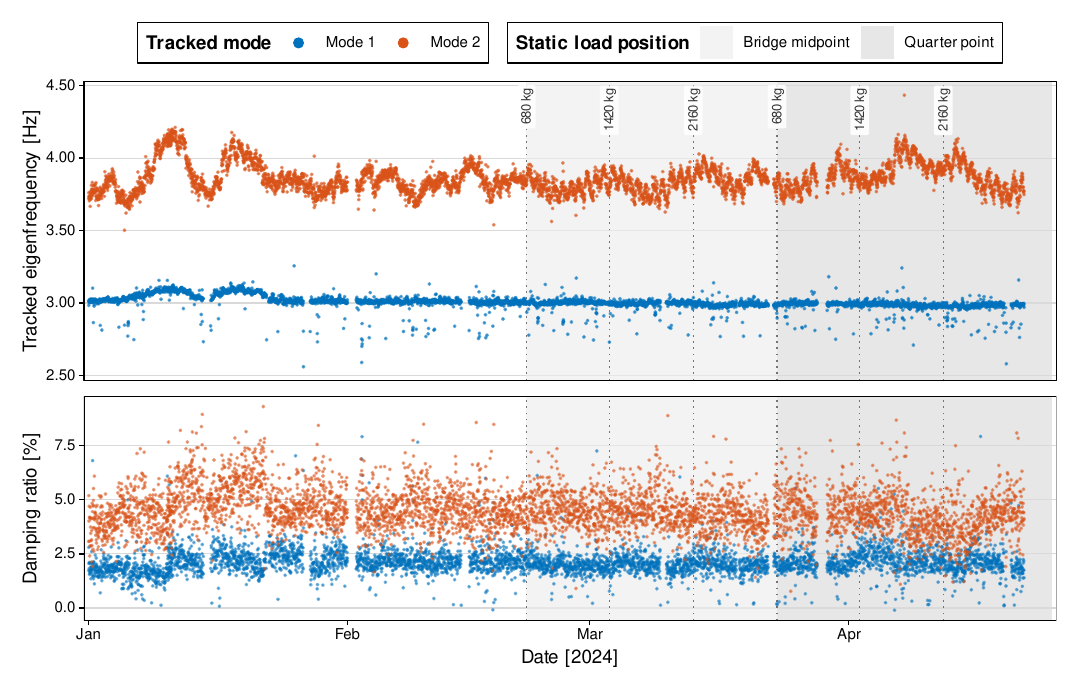}
    \caption{
            Tracked modal parameters during the long-term static load test. 
            Natural frequencies (top) and modal damping ratios (bottom) are shown for the first two modes, tracked at nominal \SI{1800}{s} intervals using a MAC threshold of 0.8. Grey bands indicate the long-term static load positions, and dotted vertical lines mark changes in added mass.
            }
    \label{fig:modal_parameters}
\end{figure}

Figure~\ref{fig:modal_parameters} shows the tracked modal parameters during the long-term static load test. 
The natural frequencies and modal damping ratios of the first two modes were tracked at nominal \SI{1800}{s} intervals using a MAC threshold of 0.8. 
The period before the additional masses were applied represents the bridge's reference state. 
During the subsequent loading periods, only small changes in the tracked natural frequencies are observed, with the clearest reduction occurring for the highest added mass of \SI{2160}{kg}. 
In contrast, the damping ratios exhibit substantially greater scatter and show no systematic change with increasing added mass. 
This result is important for interpreting the data set: the added masses represent a deliberately small perturbation, and standard modal-parameter tracking alone is only weakly sensitive to this loading scenario under environmental and operational variability. 
The long-term added-mass test should therefore be understood as a benchmark for detection-limit studies and confounder-aware monitoring methods rather than as a case with an obvious modal signature.

\begin{figure}[!htbp]
    \centering
    \includegraphics[width = \textwidth]{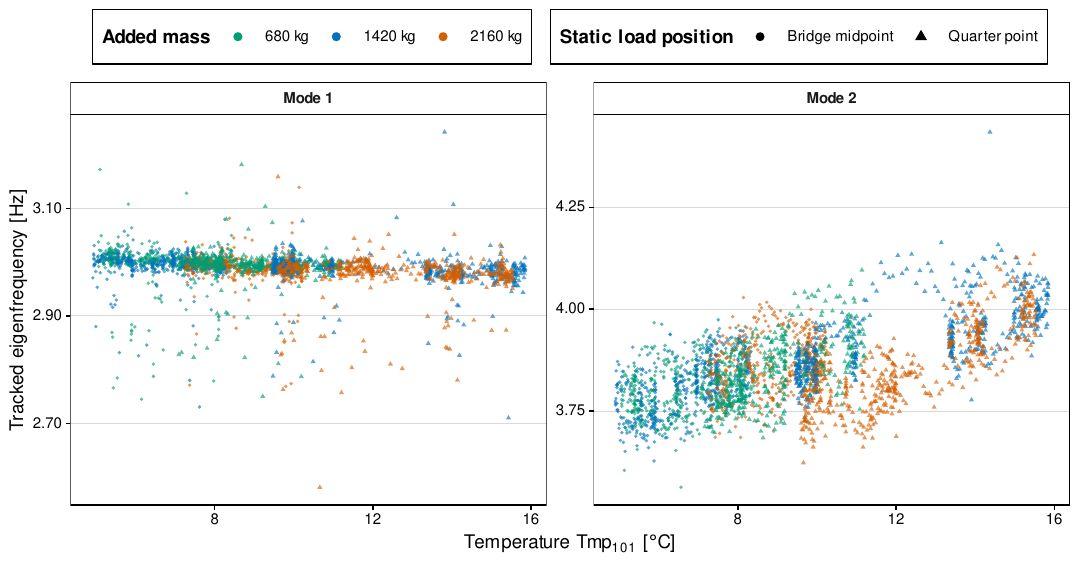}
    \caption{
            Relationship between tracked eigenfrequencies of the first two modes and temperature Tmp\_101 during the long-term static load-test period. 
            Marker color indicates the added mass, and marker shape indicates the static load position. 
            }
    \label{fig:eigenfrequency_temperature}
\end{figure}

To support this interpretation, Figure~\ref{fig:eigenfrequency_temperature} compares the tracked eigenfrequencies with the temperature measured by Tmp\_101 during the same long-term static load-test period. 
The scatter shows that the modal frequencies, especially Mode~2, vary with temperature. 
Because temperature, load stage, and static load position are not independent in this campaign, the figure is used to document environmental confounding rather than to isolate a purely mass-induced frequency shift. 
Therefore, it is not used to establish a complete seasonal temperature-frequency model.

The derived tracked eigenfrequency time series for the first two modes from January 1, 2024, to the end of the long-term static load test on April 21, 2024, are included in the Zenodo archive to support reuse of the modal-tracking results.

\FloatBarrier

%%%%%%%%%%%%%%%%%%%%%%%%%%%%%%%%%%%%%%%%%%%%%%%%
\subsection{Strain measurement}
%%%%%%%%%%%%%%%%%%%%%%%%%%%%%%%%%%%%%%%%%%%%%%%%

The strain measurements were assessed using two complementary checks. 
First, the short-term static response was evaluated for the loaded truck parked at the bridge midpoint (PP1-2), using the \SI{10}{s} interval immediately before the parking window as the baseline reference. 
Second, repeated crawl-speed passages were aligned using the laser-barrier measurements and truck geometry to compare response profiles under unloaded and loaded truck configurations. 
This positioning-based evaluation follows the emphasis on accurate load placement in short-term bridge load testing \citep{Jaelani.etal_2023, Sanio.etal_2022, Schartner.etal_2022}.

\begin{figure}[!htbp]
    \centering
    \includegraphics[width = 1\textwidth]{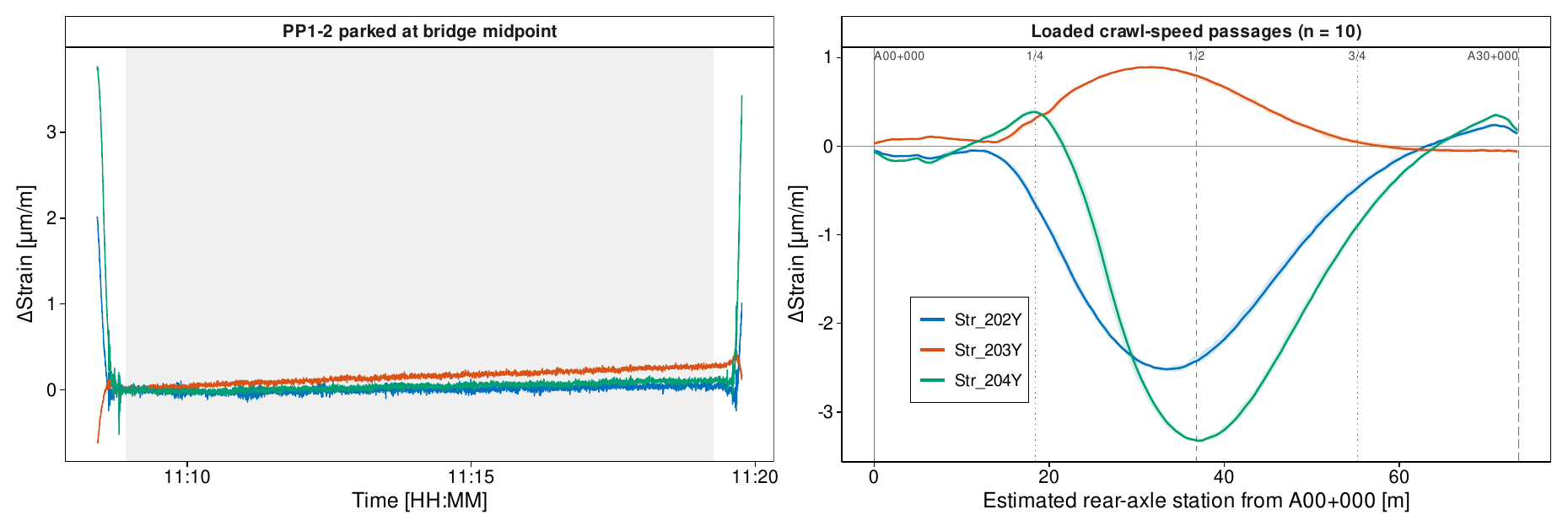}
    \caption{
            Strain response during the short-term static and dynamic load tests. 
            The left panel shows the baseline-corrected strain response while the loaded truck was parked at the bridge midpoint (PP1-2); grey shading marks the measurement window. 
            The right panel shows mean loaded crawl-speed strain profiles over 10 repeated passages, plotted against the estimated rear-axle station from A00+000 using the LS\_Anfang/LS\_Ende laser-barrier crossings and truck geometry. 
            Shaded bands indicate the 10th--90th percentile range of the run-level binned profiles and provide a repeatability check for the loaded crawl-speed passages.}
    \label{fig:strain_short_term_scenarios}
\end{figure}

Because initial strain levels differed between sensors, each signal was baseline-corrected before comparing relative strain changes across load conditions. 
Figure~\ref{fig:strain_short_term_scenarios} shows strain data from sensors Str\_202Y, Str\_203Y, and Str\_204Y, located on the web of the concrete box girder near the west abutment (see Figure~\ref{fig:Top_View}). 
The static panel shows small sensor-specific strain changes during the parking interval, while the loaded crawl-speed profiles provide a repeatability check over repeated passages under the same truck configuration.

\begin{figure}[!htbp]
    \centering
    \includegraphics[width = 1\textwidth]{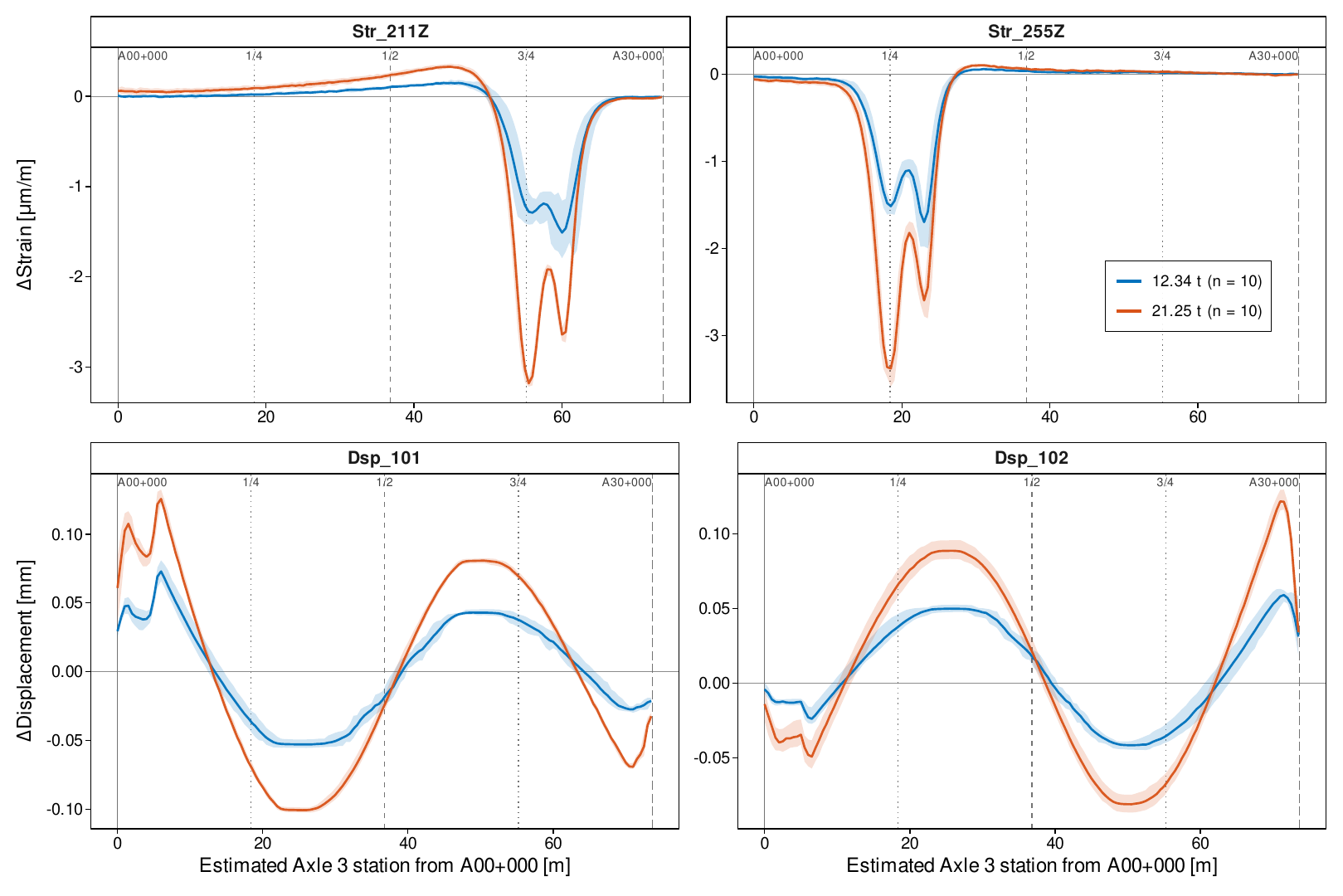}
    \caption{
            Mean crawl-speed load-response profiles for selected strain sensors (top) and displacement sensors (bottom). 
            Responses are plotted against the estimated rear-axle station (Axle~3) from A00+000, obtained from the LS\_Anfang and LS\_Ende laser-barrier crossings, bridge station length, and truck axle geometry. Solid lines show the mean over repeated runs; shaded bands indicate the 10th--90th percentile range of the run-level binned profiles. 
            The narrow bands provide a repeatability check across repeated crawl-speed passages and load configurations.
            The third unloaded crawl-speed run was excluded because the measurement started too late.
            }
    \label{fig:strain_loading_scenarios}
\end{figure}

Strain responses from selected sensors Str\_211Z and Str\_255Z, installed on the web of the concrete box girder (see Figure~\ref{fig:Top_View}), were used to compare the bridge response under the unloaded and loaded truck configurations. The crawl-speed passages were evaluated because dynamic effects are minimized under these conditions. For each load configuration, the repeated passages were transformed to an estimated rear-axle station using the laser-barrier measurements and truck geometry, and then aggregated into mean response profiles with uncertainty bands (Figure~\ref{fig:strain_loading_scenarios} (top). The resulting profiles show good repeatability between repeated crossings and the expected increase in strain magnitude for the loaded truck configuration (\SI{21250}{kg}) compared with the unloaded configuration (\SI{12340}{kg}).

For the strain and displacement response-profile figures, baseline-corrected signals were smoothed using a Savitzky--Golay filter with polynomial order 3 and window length 21 samples before run-level binning \citep{Schafer_2011}.

The strain response profiles in Figures~\ref{fig:strain_short_term_scenarios} and~\ref{fig:strain_loading_scenarios} are influence-line response profiles of the bridge under the moving truck load; per-axle influence lines can be extracted from them, demonstrating the influence-line--based assessment.

\begin{figure}[!htbp]
    \centering
    \includegraphics[width=1\textwidth]{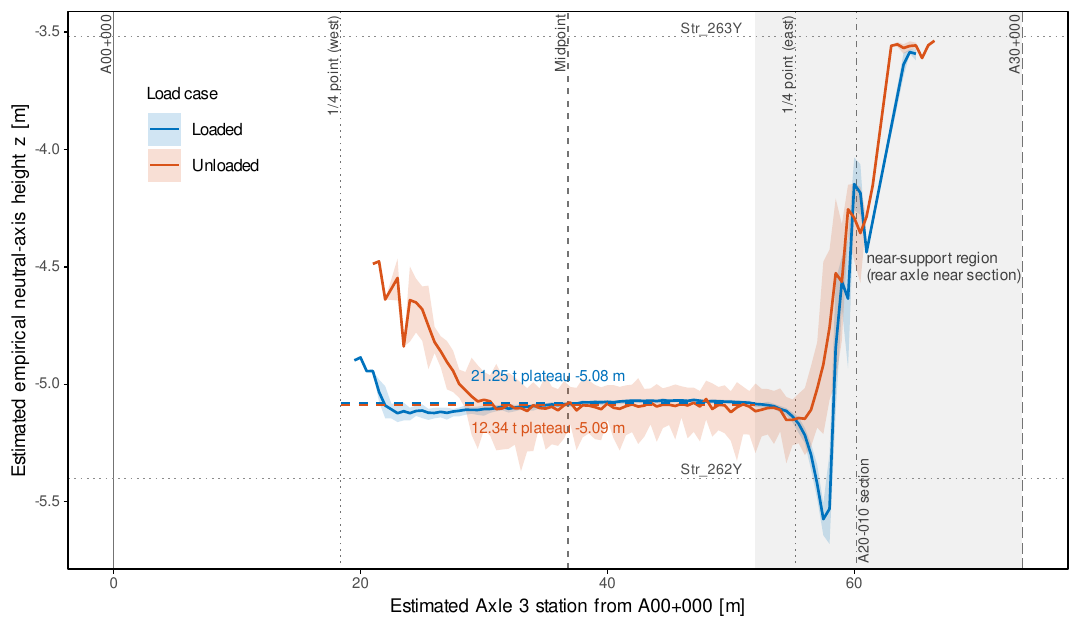}
     \caption{Estimated empirical neutral-axis height $z_{\mathrm{NA}}$ at section A20-010 from the strain pair Str\_262Y ($z = \SI{-5.40}{m}$) and Str\_263Y ($z = \SI{-3.52}{m}$), with the section location shown in Figure~\ref{fig:Top_View} and the sensor coordinates listed in Table~\ref{tab:long}, plotted against the estimated Axle~3 station from A00+000 across the loaded (\SI{21.25}{t}, ten runs) and unloaded (\SI{12.34}{t}, ten runs) crawl-speed crossings. Solid lines: per-load-case mean over runs per $\SI{0.5}{m}$ bin; shaded bands: 10th--90th percentile across runs of the same load case; dashed horizontal segments: plateau median over the plateau window for each load case. Horizontal dotted lines mark the two sensor heights; the dot-dashed marker indicates the A20-010 section station; the grey band marks the near-support region where the rear axle is over or past the section, and the two-sensor estimate departs from the plateau.}
    \label{fig:neutral_axis}
\end{figure}

As a worked example of the neutral-axis reuse case, the live-load strains at two heights of section A20-010 (Str\_262Y at $z = \SI{-5.40}{m}$ and Str\_263Y at $z = \SI{-3.52}{m}$, lever arm $\SI{1.88}{m}$; see Figure~\ref{fig:Top_View} for the section location and Table~\ref{tab:long} for the sensor coordinates) were used to estimate the neutral-axis height during the ten loaded crawl-speed crossings, assuming a linear strain distribution over the depth, $\varepsilon(z) = \varepsilon_0 + \kappa\,z$, so that the neutral-axis height follows from $z_{\mathrm{NA}} = z_1 - \Delta\varepsilon_1\,(z_2 - z_1)/(\Delta\varepsilon_2 - \Delta\varepsilon_1)$ with baseline-subtracted strains $\Delta\varepsilon_i$ (Figure~\ref{fig:neutral_axis}).
The baseline is the median of the pre-entry samples before the first \texttt{LS\_Anfang} axle crossing, removing the static self-weight, prestress, thermal, and sensor-zero offsets that would otherwise contaminate the zero-crossing estimate. The resulting $z_{\mathrm{NA}}$ is therefore the empirical, strain-derived height at which the live-load bending strain vanishes for the two-sensor pair under the applied truck loading; it is reported as such, without reference to a theoretical section centroid, which would require independent geometric properties of the box girder cross-section at A20-010 and is treated as a separate exercise.
While the rear axle loads the instrumented section, the estimated neutral-axis height settles to a stable value of about $\SI{-5.08}{m}$ for the loaded truck (\SI{21.25}{t}) and $\SI{-5.09}{m}$ for the unloaded truck (\SI{12.34}{t}) (median across the $18.4$ to $\SI{52}{m}$ Axle~3-station plateau window), and the estimate is repeatable across the crossings of each load configuration. The agreement to within $\SI{0.01}{m}$ between the two load levels indicates that the empirical neutral-axis position is independent of the applied load magnitude over this range, as expected for an elastic, undamaged section. This demonstrates that the data set supports neutral-axis extraction.

\FloatBarrier

%%%%%%%%%%%%%%%%%%%%%%%%%%%%%%%%%%%%%%%%%%%%%%%%
\subsection{Displacement measurement}
%%%%%%%%%%%%%%%%%%%%%%%%%%%%%%%%%%%%%%%%%%%%%%%%

Three independent observations of the displacement response are presented: static truck loading at the quarter spans, dynamic crawl-speed crossings, and long-term thermal coupling.
Displacement sensors are grouped by location: Dsp\_101 and Dsp\_103 at the western expansion joints, Dsp\_102 and Dsp\_104 at the eastern expansion joints.

\begin{figure}[!htbp]
    \centering
    \includegraphics[width = 1\textwidth]{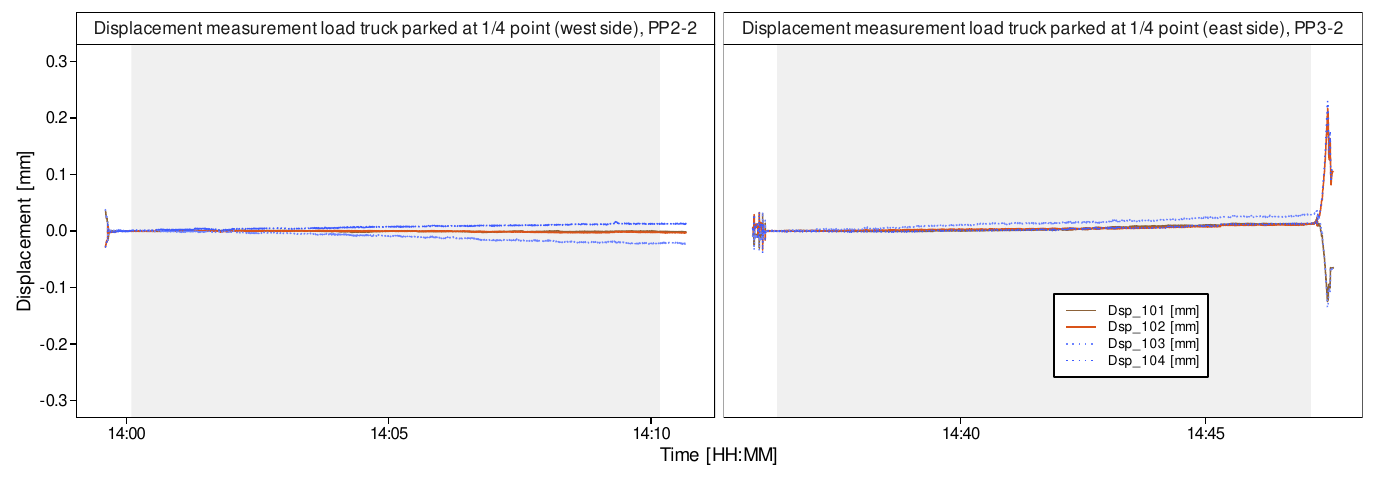}
    \caption{
            Baseline-corrected displacement measurements at the expansion joints while the load truck was stationary at PP2-2 and PP3-2.
            Grey shading marks the stationary parking intervals; displacements were zero-referenced using the median of the 10~s interval immediately before each parking window.
            }
    \label{fig:displacement_st}
\end{figure}

Figure~\ref{fig:displacement_st} shows the displacement curves when the \SI{21250}{kg} truck was stationary at the quarter span on both bridge sections (PP2-2 and PP3-2).
The median load-induced displacement during the parking intervals, baseline-subtracted using the $\SI{10}{s}$ pre-test window and averaged across both parking panels and the two sensors per joint side, was $\SI{0.004}{mm}$ at the eastern joints (Dsp\_102, Dsp\_104) and $\SI{0.006}{mm}$ at the western joints (Dsp\_101, Dsp\_103); the joints respond at the micrometer scale, consistent with the truck being parked at the quarter spans rather than at either bridge end.

\begin{figure}[!htbp]
    \centering
    \includegraphics[width=0.90\textwidth]{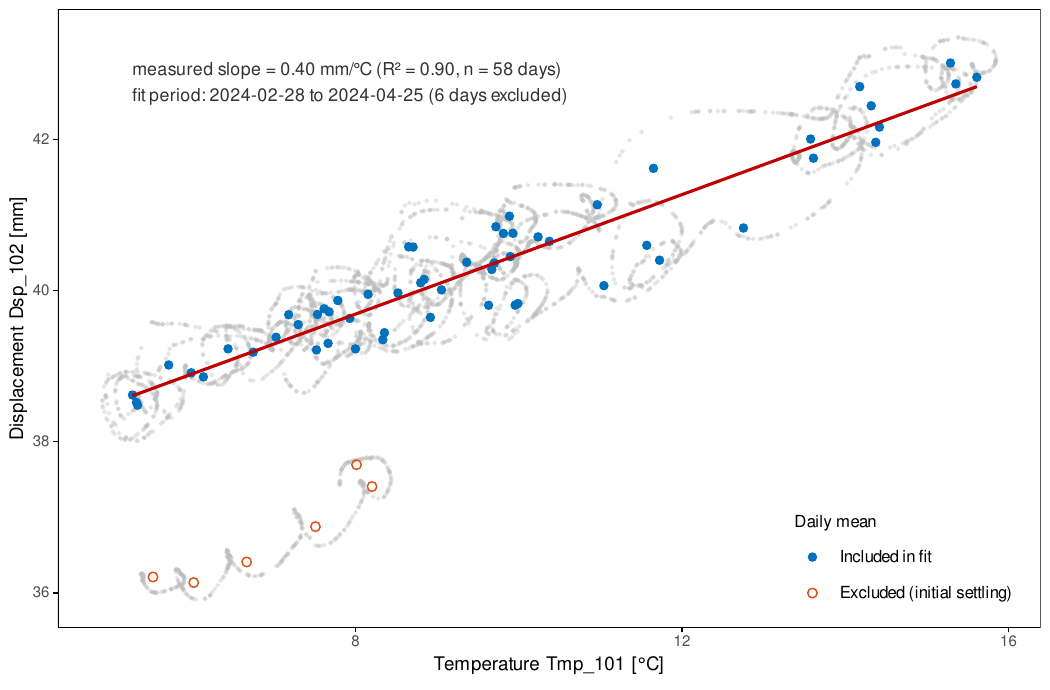}
    \caption{Eastern-joint displacement Dsp\_102 versus embedded structure temperature Tmp\_101 over the long-term static-load-test period.
    Grey points: sub-daily raw samples (thinned for plotting) showing the intra-day hysteresis between the point temperature sensor and the bulk structure.
    Filled blue circles: daily means included in the linear fit.
    Open orange circles: daily means from the initial settling period excluded from the fit.
    Red line: linear fit on the included daily means.
    The in-plot annotation (upper left) reports the measured slope, coefficient of determination, and the fit period; the legend (lower right) identifies the two daily-mean classes. Note that an increase in the displacement represents an extension of the bridge.}
    \label{fig:temp_disp_validation}
\end{figure}

To characterize the long-term coupling between temperature and displacement at the pavement transition structures, the relationship between eastern-joint displacement Dsp\_102 and the embedded structure temperature Tmp\_101 over the long-term static-load-test period is shown in Figure~\ref{fig:temp_disp_validation}.

For the loaded (\SI{21250}{kg}) and unloaded (\SI{12340}{kg}) crawl-speed crossings, the displacement profiles of Dsp\_101 and Dsp\_102 (displacement panel of Figure~\ref{fig:strain_loading_scenarios}) behave as expected: the heavier configuration produced larger displacements, and both sensors responded consistently.
Across the dynamic crossings, the maximum horizontal displacements under the loaded truck were $\SI{0.13}{mm}$ (west, Dsp\_101) and $\SI{0.12}{mm}$ (east, Dsp\_102), taken from the run-mean profiles over the ten loaded crawl-speed crossings.
Dsp\_101 and Dsp\_102 are shown as representative western and eastern joints.

It is imperative to specify the measurement direction to ensure that an observed increase in displacement accurately reflects the bridge's actual expansion. The first six days (2024-02-22 to 2024-02-27) are systematically about $\SI{2}{mm}$ below the rest of the series with no matching temperature excursion.
These days are excluded from the linear fit; the fit therefore covers 2024-02-28 to 2024-04-25 (58 daily means, $R^2 = 0.90$, slope $\SI{0.40}{mm/\celsius}$).
Dsp\_102 is shown as a representative eastern joint.

Together, these three modes confirm that the displacement sensors respond consistently with applied load and with ambient temperature over the campaign, supporting their use for both short-term load tests and long-term thermal-response characterization.

\FloatBarrier

%%%%%%%%%%%%%%%%%%%%%%%%%%%%%%%%%%%%%%%%%%%%%%%%
\section{Processing of Measurement Data}
%%%%%%%%%%%%%%%%%%%%%%%%%%%%%%%%%%%%%%%%%%%%%%%%

The measurement data were converted from the proprietary Dewesoft format (.dxd) to the open-source HDF5 format, which supports metadata and is widely supported across analysis software. 
During measurement, the recorded quantities are converted to target variables using the manufacturer's conversion factors. 
All sensor data were transmitted via cable for higher transmission security. 
For the long-term static load tests, the published data were aggregated to \SI{1}{Hz} so that the full multi-week duration could be provided in a single, manageable record; the dynamic load test data were not downsampled. In addition to the direct measurement data, the archive includes derived tracked eigenfrequency time series for the first two modes from January 1, 2024, to the end of the long-term static load test on April 21, 2024. The complete raw data (approximately \SI{7.7}{GB} per day) and the full reference-state monitoring data are available from the corresponding author upon reasonable request, subject to a cooperation agreement required by the infrastructure operator.

%%%%%%%%%%%%%%%%%%%%%%%%%%%%%%%%%%%%%%%%%%%%%%%%
\section{Data Format and Hierarchical Structure}
%%%%%%%%%%%%%%%%%%%%%%%%%%%%%%%%%%%%%%%%%%%%%%%%

\begin{figure}[!htb]
    \centering
    \includegraphics[width = .85\textwidth]{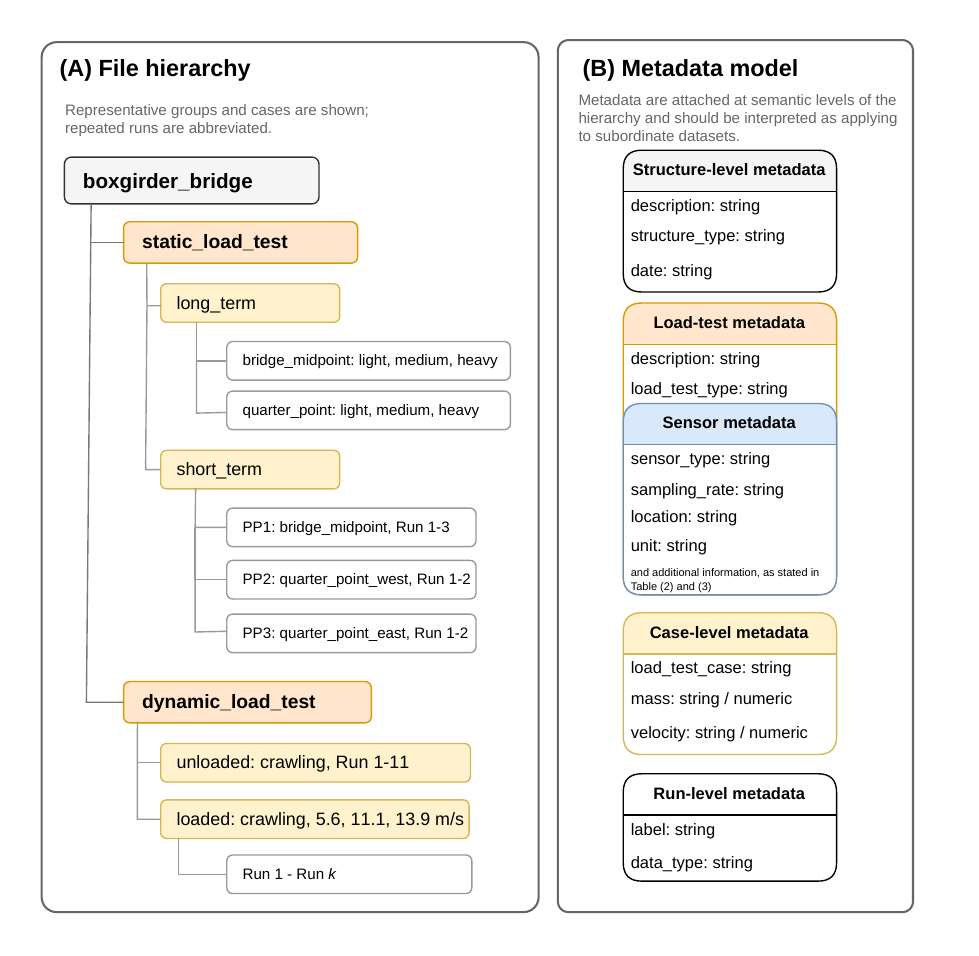}
    \caption{Hierarchical organization of the archived data of the load tests and associated metadata.}
    \label{fig:file_structure}
\end{figure}

All measurements were stored in HDF5 files and archived in a zip file with various directories. 
The structure of the zip file is shown in Figure \ref{fig:file_structure}. 
The files are structured as follows. First, a distinction is made between the type of load test, i.e., whether it is a static or dynamic load. 
Then, the corresponding load case is differentiated, for example, which weight was applied at which position. 
All sensors were stored in a single file to simplify data use. 
Derived modal-tracking results, including the tracked eigenfrequencies used in Figure~\ref{fig:modal_parameters}, are provided as separate analysis-ready data products in the archive.

%%%%%%%%%%%%%%%%%%%%%%%%%%%%%%%%%%%%%%%%%%%%%%%%
\section{Conclusion}
%%%%%%%%%%%%%%%%%%%%%%%%%%%%%%%%%%%%%%%%%%%%%%%%

Following the description of the concrete box girder bridge and the measurement system, the static and dynamic load tests were presented. 
In addition, the open-access database, the processing of measurement data, and the file structure were presented to enable further use of the collected measurement data. 
The static load tests were conducted over a period of multiple weeks. 
By gradually increasing the applied masses, it is possible to assess the minimum detectable mass change, which is particularly useful for damage detection methods that account for temperature variations, such as those described in \citep{Wittenberg.etal_2025, Neumann.etal_2025}.
Dynamic crossings were repeated multiple times, enabling repeatability assessment and uncertainty quantification of the response profiles.
The different speeds of the dynamic load tests enable a detailed analysis of the influence of the speed of the crossing on the bridge behavior.
The short-term static and dynamic load tests were conducted within a single day to ensure comparable boundary conditions and to minimize environmental differences between the two test phases.

In summary, the load tests were successfully conducted and provide a solid foundation for further research. 
Integrating SHM with load tests offers a valuable approach for validating SHM data and bridge design assumptions. 
Given the quantity and thorough documentation of the measurements, the data can be used to develop and validate algorithms that correlate various measured parameters. The influence-line response profiles are demonstrated here (Figures~\ref{fig:strain_short_term_scenarios} and~\ref{fig:strain_loading_scenarios}) \citep{Zheng.etal_2021}, and the data further support neutral-axis extraction \citep{Sigurdardottir.etal_2013} and the reconstruction of signals and features under varying environmental conditions \citep{Neumann.etal_2026a}.

The data set has several limitations. 
Because the bridge remains in service, only non-destructive load tests were permitted; test configurations that could risk short- or long-term changes to the structural condition were excluded. The data, therefore, represent controlled service-level loading rather than ultimate or damage-inducing behavior. 
The bridge name and location are anonymized per the infrastructure operator's requirements. 
This restriction limits independent site inspection, external cross-referencing with inspection reports, and future visits by third parties, but it was necessary to make the load-test data publicly available. 
The long-term added masses produced only small changes in the tracked modal parameters, which limits the direct use of simple frequency- or damping-threshold methods and motivates the use of analysis methods that explicitly account for environmental and operational variability. 
Finally, the short-term static and dynamic load tests were carried out on a single day, so the live-load response was captured at essentially one temperature; repeating these tests across different temperatures would enable a more direct assessment of temperature effects on the load response and is left for future work.

%%%%%%%%%%%%%%%%%%%%%%%%%%%%%%%%%%%%%%%%%%%%%%%%
\section{Data availability statement}
%%%%%%%%%%%%%%%%%%%%%%%%%%%%%%%%%%%%%%%%%%%%%%%%

The load test data set is available on Zenodo~\citep{koehncke_2026}. 
The archive contains the HDF5 load-test data, associated metadata and documentation, and derived tracked eigenfrequency time series for the first two modes from January 1, 2024, to the end of the long-term static load test on April 21, 2024.
Further data of the reference state of this study and additional supporting documentation, materials, and plans are available from the corresponding author upon reasonable request.
Zenodo is an open data repository for preserving research data. The datasets can be cited using the DOI. 

%%%%%%%%%%%%%%%%%%%%%%%%%%%%%%%%%%%%%%%%%%%%%%%%
\section{Acknowledgements}
%%%%%%%%%%%%%%%%%%%%%%%%%%%%%%%%%%%%%%%%%%%%%%%%

This research is funded by dtec.bw -- Digitalization and Technology Research Center of the Bundeswehr. 
dtec.bw is funded by the European Union -- NextGenerationEU. We thank all dtec.bw Project supporters, ATS Alpha Tech Services GmbH, iseatec GmbH, REVOTEC zt GmbH, and the Federal Logistics and Mobility Office (BALM) for providing and weighing the test truck. We also thank the two anonymous reviewers, the Associate Editor, and the Editor for their constructive comments and guidance. We thank Dr.-Ing. Jan Grashorn for his support in preparing the data set for publication on Zenodo.

\bibliographystyle{plainnat}
\bibliography{literature}

%%%%%%%%%%%%%%%%%%%%%%%%%%%%%%%%%%%%%%%%%%%%%%%%
\appendix
\section{Appendix}
%%%%%%%%%%%%%%%%%%%%%%%%%%%%%%%%%%%%%%%%%%%%%%%%

\begin{longtable}{
    @{}
    L{1.65cm}
    L{1.65cm}
    L{2.85cm}
    S[table-format=-1.2]
    S[table-format=-2.2]
    S[table-format=-1.2]
    O{4.50cm}
    @{}
}
\caption{Sensor name, location, and coordinate system.}
\label{tab:long} \\

\toprule
\multicolumn{1}{c}{\textbf{Type}} &
\multicolumn{1}{c}{\textbf{Lateral axis}} &
\multicolumn{1}{c}{\textbf{Position}} &
{\textbf{X [m]}} &
{\textbf{Y [m]}} &
{\textbf{Z [m]}} &
\multicolumn{1}{c}{\textbf{Orientation}} \\
\midrule
\endfirsthead

\multicolumn{7}{c}%
{{\bfseries \tablename\ \thetable{} -- continued from previous page}} \\
\toprule
\multicolumn{1}{c}{\textbf{Type}} &
\multicolumn{1}{c}{\textbf{Lateral axis}} &
\multicolumn{1}{c}{\textbf{Position}} &
{\textbf{X [m]}} &
{\textbf{Y [m]}} &
{\textbf{Z [m]}} &
\multicolumn{1}{c}{\textbf{Orientation}} \\
\midrule
\endhead

\midrule
\multicolumn{7}{r}{{Continued on next page}} \\
\endfoot

\bottomrule
\endlastfoot

Acc\_301X & A10+063 & box girder & 1.27 & 18.68 & -1.31 & 1 \\
Acc\_301Y & A10+063 & box girder & 1.27 & 18.68 & -1.31 & -1 \\
Acc\_301Z & A10+063 & box girder & 1.27 & 18.68 & -1.31 & -1 \\
Acc\_302X & A10+125 & box girder & 1.27 & 25.07 & -1.63 & 1 \\
Acc\_302Y & A10+125 & box girder & 1.27 & 25.07 & -1.63 & -1 \\
Acc\_302Z & A10+125 & box girder & 1.27 & 25.07 & -1.63 & -1 \\
Acc\_303X & A10+180 & box girder & 1.27 & 30.27 & -1.90 & 1 \\
Acc\_303Y & A10+180 & box girder & 1.27 & 30.27 & -1.90 & -1 \\
Acc\_303Z & A10+180 & box girder & 1.27 & 30.27 & -1.90 & -1 \\
Acc\_304X & A20-180 & box girder & 1.27 & 44.26 & -2.62 & -1 \\
Acc\_304Y & A20-180 & box girder & 1.27 & 44.26 & -2.62 & 1 \\
Acc\_304Z & A20-180 & box girder & 1.27 & 44.26 & -2.62 & -1 \\
Acc\_305X & A20-125 & box girder & 1.27 & 49.81 & -2.90 & -1 \\
Acc\_305Y & A20-125 & box girder & 1.27 & 49.81 & -2.90 & 1 \\
Acc\_305Z & A20-125 & box girder & 1.27 & 49.81 & -2.90 & -1 \\
Acc\_306X & A20-063 & box girder & 1.27 & 56.19 & -3.23 & -1 \\
Acc\_306Y & A20-063 & box girder & 1.27 & 56.19 & -3.23 & 1 \\
Acc\_306Z & A20-063 & box girder & 1.27 & 56.19 & -3.23 & -1 \\
Acc\_351X & A10+063 & box girder & -1.27 & 17.90 & -1.32 & 1 \\
Acc\_351Y & A10+063 & box girder & -1.27 & 17.90 & -1.32 & -1 \\
Acc\_351Z & A10+063 & box girder & -1.27 & 17.90 & -1.32 & -1 \\
Acc\_352X & A10+125 & box girder & -1.27 & 24.26 & -1.64 & 1 \\
Acc\_352Y & A10+125 & box girder & -1.27 & 24.26 & -1.64 & -1 \\
Acc\_352Z & A10+125 & box girder & -1.27 & 24.26 & -1.64 & -1 \\
Acc\_353X & A10+180 & box girder & -1.27 & 29.45 & -1.91 & 1 \\
Acc\_353Y & A10+180 & box girder & -1.27 & 29.45 & -1.91 & -1 \\
Acc\_353Z & A10+180 & box girder & -1.27 & 29.45 & -1.91 & -1 \\
Acc\_354X & A20-180 & box girder & -1.27 & 43.40 & -2.63 & -1 \\
Acc\_354Y & A20-180 & box girder & -1.27 & 43.40 & -2.63 & 1 \\
Acc\_354Z & A20-180 & box girder & -1.27 & 43.40 & -2.63 & -1 \\
Acc\_355X & A20-125 & box girder & -1.27 & 49.02 & -2.91 & -1 \\
Acc\_355Y & A20-125 & box girder & -1.27 & 49.02 & -2.91 & 1 \\
Acc\_355Z & A20-125 & box girder & -1.27 & 49.02 & -2.91 & -1 \\
Acc\_356X & A20-063 & box girder & -1.27 & 55.40 & -3.24 & 1 \\
Acc\_356Y & A20-063 & box girder & -1.27 & 55.40 & -3.24 & -1 \\
Acc\_356Z & A20-063 & box girder & -1.27 & 55.40 & -3.24 & -1 \\

\midrule
Inc\_201 & A10+010 & box girder & 0.07 & 13.06 & -0.93 & X = +1, Y = -1, Z = -1 \\
Inc\_202 & A10+180 & box girder & 0.02 & 30.04 & -1.84 & X = +1, Y = -1, Z = -1 \\
Inc\_203 & A20-180 & box girder & 0.01 & 44.03 & -2.56 & X = -1, Y = +1, Z = -1 \\
Inc\_204 & A20-010 & box girder & 0.01 & 61.03 & -3.43 & X = +1, Y = -1, Z = -1 \\

\midrule
Str\_201X & A10+000 & support & 2.04 & 11.69 & -4.75 & - \\
Str\_201Z & A10+000 & support & 2.04 & 11.85 & -4.59 & - \\
Str\_202Z & A10+010 & box girder & 1.95 & 13.74 & -2.56 & - \\
Str\_202Y & A10+010 & box girder & 1.95 & 13.74 & -2.69 & - \\
Str\_203Z & A10+010 & box girder & 1.95 & 13.66 & -1.26 & - \\
Str\_203Y & A10+010 & box girder & 1.95 & 13.66 & -1.15 & - \\
Str\_204Z & A10+063 & box girder & 1.95 & 19.10 & -2.40 & - \\
Str\_204Y & A10+063 & box girder & 1.95 & 19.10 & -2.45 & - \\
Str\_205Z & A10+063 & box girder & 1.95 & 19.10 & -1.55 & - \\
Str\_205Y & A10+063 & box girder & 1.95 & 19.10 & -1.43 & - \\
Str\_206Z & A10+180 & box girder & 1.95 & 30.91 & -2.25 & - \\
Str\_206Y & A10+180 & box girder & 1.95 & 31.02 & -2.32 & - \\
Str\_207Z & A10+180 & box girder & 1.94 & 31.08 & -2.15 & - \\
Str\_207Y & A10+180 & box girder & 1.94 & 31.08 & -2.04 & - \\
Str\_208Z & A20-180 & box girder & 1.94 & 44.51 & -3.04 & - \\
Str\_208Y & A20-180 & box girder & 1.94 & 44.51 & -3.16 & - \\
Str\_209Z & A20-180 & box girder & 1.94 & 44.38 & -2.83 & - \\
Str\_209Y & A20-180 & box girder & 1.94 & 44.51 & -2.73 & - \\
Str\_210Z & A20-063 & box girder & 1.94 & 56.24 & -4.49 & - \\
Str\_210Y & A20-063 & box girder & 1.94 & 56.24 & -4.64 & - \\
Str\_211Z & A20-063 & box girder & 1.94 & 56.34 & -3.44 & - \\
Str\_211Y & A20-063 & box girder & 1.94 & 56.34 & -3.34 & - \\
Str\_212Z & A20-010 & box girder & 1.94 & 61.67 & -5.05 & - \\
Str\_212Y & A20-010 & box girder & 1.94 & 61.72 & -5.20 & - \\
Str\_213Z & A20-010 & box girder & 1.94 & 61.72 & -3.79 & - \\
Str\_213Y & A20-010 & box girder & 1.94 & 61.72 & -3.65 & - \\
Str\_214X & A20-000 & support & 1.98 & 63.45 & -7.23 & - \\
Str\_214Z & A20-000 & support & 1.98 & 63.58 & -7.39 & - \\
Str\_251X & A10+000 & support & -1.91 & 10.60 & -4.75 & - \\
Str\_251Z & A10+000 & support & -1.91 & 10.65 & -4.59 & - \\
Str\_252Z & A10+010 & box girder & -1.92 & 12.38 & -2.70 & - \\
Str\_252Y & A10+010 & box girder & -1.92 & 12.38 & -2.80 & - \\
Str\_253Z & A10+010 & box girder & -1.92 & 12.38 & -1.29 & - \\
Str\_253Y & A10+010 & box girder & -1.92 & 12.38 & -1.18 & - \\
Str\_254Z & A10+063 & box girder & -1.93 & 17.78 & -2.55 & - \\
Str\_254Y & A10+063 & box girder & -1.93 & 17.82 & -2.46 & - \\
Str\_255Z & A10+063 & box girder & -1.93 & 17.78 & -1.55 & - \\
Str\_255Y & A10+063 & box girder & -1.93 & 17.82 & -1.46 & - \\
Str\_256Z & A10+180 & box girder & -1.94 & 29.97 & -2.29 & - \\
Str\_256Y & A10+180 & box girder & -1.94 & 29.90 & -2.40 & - \\
Str\_257Z & A10+180 & box girder & -1.94 & 29.83 & -2.18 & - \\
Str\_257Y & A10+180 & box girder & -1.94 & 29.83 & -2.08 & - \\
Str\_258X & A20-180 & box girder & -1.32 & 42.72 & -2.97 & - \\
Str\_258Y & A20-180 & box girder & -1.37 & 42.86 & -2.98 & - \\
Str\_259X & A20-180 & box girder & -1.44 & 42.72 & -2.56 & - \\
Str\_259Y & A20-180 & box girder & -1.39 & 42.87 & -2.56 & - \\
Str\_260X & A20-063 & box girder & -1.72 & 55.24 & -4.67 & - \\
Str\_260Y & A20-063 & box girder & -1.82 & 55.14 & -4.64 & - \\
Str\_261X & A20-063 & box girder & -1.47 & 54.95 & -3.20 & - \\
Str\_261Y & A20-063 & box girder & -1.38 & 54.90 & -3.17 & - \\
Str\_262X & A20-010 & box girder & -1.39 & 60.32 & -5.41 & - \\
Str\_262Y & A20-010 & box girder & -1.36 & 60.22 & -5.40 & - \\
Str\_263X & A20-010 & box girder & -1.79 & 60.19 & -3.55 & - \\
Str\_263Y & A20-010 & box girder & -1.68 & 60.19 & -3.52 & - \\
Str\_264X & A20-000 & support & -1.90 & 62.23 & -7.21 & - \\
Str\_264Z & A20-000 & support & -1.90 & 62.39 & -7.45 & - \\

\midrule
Dsp\_101 & A00-000 & road & 4.92 & 0.34 & -0.01 & - \\
Dsp\_102 & A30+000 & road & 4.92 & 73.98 & -3.78 & - \\
Dsp\_103 & A00-000 & road & -4.92 & -1.44 & -0.11 & - \\
Dsp\_104 & A30+000 & road & -4.92 & 72.25 & -3.88 & - \\

\midrule
Tmp\_101 & A10+010 & support & 1.90 & 11.93 & -4.30 & - \\
Tmp\_102 & A10+063 & box girder & 1.95 & 18.98 & -2.68 & - \\
Tmp\_103 & A10+063 & box girder & 1.95 & 18.98 & -1.43 & - \\
Tmp\_104 & A10+125 & box girder & 1.95 & 30.68 & -2.35 & - \\
Tmp\_105 & A10+125 & box girder & 1.95 & 30.68 & -2.04 & - \\
Tmp\_106 & A20-063 & box girder & 1.95 & 56.38 & -4.73 & - \\
Tmp\_107 & A20-063 & box girder & 1.94 & 56.48 & -3.37 & - \\
Tmp\_108 & A20-000 & support & 1.89 & 63.27 & -7.06 & - \\
Tmp\_151 & A10+010 & support & -1.88 & 10.75 & -4.41 & - \\
Tmp\_152 & A10+063 & box girder & -1.95 & 17.69 & -2.70 & - \\
Tmp\_153 & A10+063 & box girder & -1.93 & 17.70 & -1.48 & - \\
Tmp\_154 & A10+125 & box girder & -1.95 & 29.39 & -2.39 & - \\
Tmp\_155 & A10+125 & box girder & -1.94 & 29.39 & -2.09 & - \\
Tmp\_156 & A20-063 & box girder & -1.95 & 55.09 & -4.55 & - \\
Tmp\_157 & A20-063 & box girder & -1.93 & 55.09 & -3.38 & - \\
Tmp\_158 & A20-000 & support & -1.89 & 62.12 & -6.90 & - \\

\midrule
IATmp\_101 & A10+010 & box girder & -0.90 & 13.63 & -1.06 & - \\
IATmp\_102 & A20-010 & box girder & 0.09 & 61.03 & -3.45 & - \\
Ast\_101 & A10+125 & road & -0.06 & 12.89 & -0.78 & - \\
Ast\_102 & A20-125 & road & 0.19 & 61.62 & -3.33 & - \\
Hmd\_101 & A10-010 & box girder & -0.90 & 13.63 & -1.06 & - \\
Hmd\_102 & A20-010 & box girder & -0.90 & 61.03 & -3.40 & - \\
Wst\_101 & A00-000 & road & -5.30 & -0.03 & 2.71 & - \\
Wst\_102 & A10+010 & building ground & -0.65 & 13.98 & -5.33 & - \\

\end{longtable}

\begin{table}
    \centering
    \caption{Loads of truck per state.}
    \begin{tabular}{
        l
        >{\color{black}}S[table-format = 5.0]
        >{\color{black}}S[table-format = 5.0]
    }
        \toprule
        \multicolumn{1}{c}{\textbf{Axle}} &
        {\textbf{\color{black}Empty weight [kg]}} &
        {\textbf{\color{black}Loaded weight [kg]}} \\
        \midrule
        A            & 6860  & 9340  \\
        B            & 2820  & 7160  \\
        C            & 2660  & 4750  \\
        Total weight & 12340 & 21250 \\
        \bottomrule
    \end{tabular}
    \label{tab:truck_loadings}
\end{table}
        
\begin{table}
    \centering
    \tabcolsep10.1pt 
    \small
    \caption{Overview of the times of parking positions (in UTC).}
    \begin{tabular}{l *{4}{c}}
	\toprule
        \multicolumn{1}{c}{\textbf{Position}} & 
        \multicolumn{1}{c}{\textbf{Run}} &  
        \multicolumn{1}{c}{\textbf{Start}} & 
	\multicolumn{1}{c}{\textbf{End}}& 
	\multicolumn{1}{c}{\textbf{Details}}\\
	\midrule
        PP1 & 1 & 25.04.2024 10:54.14 & 25.04.2024 11:04.45 & Bridge midpoint\\
        PP1 & 2 & 25.04.2024 11:08.55 & 25.04.2024 11:19.16 & Bridge midpoint \\
        PP1 & 3 & 25.04.2024 11:22.12 & 25.04.2024 11:32.48 & Bridge midpoint \\
        PP2 & 1 & 25.04.2024 11:43.44 & 25.04.2024 11:54.20 & Quarter point (west side) \\
        PP2 & 2 & 25.04.2024 12:00.06 & 25.04.2024 12:10.10 & Quarter point (west side) \\
        PP3 & 1 & 25.04.2024 12:23.00 & 25.04.2024 12:33.16 & Quarter point (east side) \\
        PP3 & 2 & 25.04.2024 12:36.15 & 25.04.2024 12:47.09 & Quarter point (east side) \\
	\bottomrule
    \end{tabular}
    \label{tab:parkpos}
\end{table}

\end{document}